\documentclass[%
 reprint,
 amsmath,amssymb,
 aps,
floatfix,
]{revtex4-2}

\usepackage{graphicx}
\usepackage{dcolumn}
\usepackage{bm}

\usepackage[utf8x]{inputenc}
\usepackage{xcolor}
\usepackage{braket}
\usepackage{cleveref}
\usepackage{verbatim}
\usepackage[caption=false]{subfig}
\usepackage{graphicx}
\begin{document}

\preprint{APS/123-QED}

\title{Spin-state estimation using the Stern-Gerlach experiment}

\author{Javier Martínez-Cifuentes}
\email{ajmartinezc@unal.edu.co} 
\author{K. M. Fonseca-Romero}%
\email{kmfonsecar@unal.edu.co}
\affiliation{%
 Departamento de Física, Universidad Nacional de Colombia - Sede Bogotá, Facultad de Ciencias, Grupo de Óptica e Información Cuántica, Carrera 30 Calle 45-03, C.P. 111321, Bogotá, Colombia
}%


\date{\today}

\begin{abstract}
We present a state estimation scheme for spins, using a modified setup of the Stern-Gerlach experiment, in which a beam of neutral spin-1/2 point particles interacts with a quadrupolar magnetic field.
The proposed estimation procedures, based either on a quadrant or a continuous intensity distribution detection, require a suitable initial spatial state of the beam.
The statistical characterization of the estimators of the initial spin state allows us not only to associate an error to the estimated parameters, but also to define a measure for comparing estimation procedures corresponding to different Stern-Gerlach setups.
\end{abstract}

\maketitle


\section{\label{sec:introduction}Introduction}
\begin{figure*}[htbp!]
\centering
\includegraphics[width=0.98\textwidth]{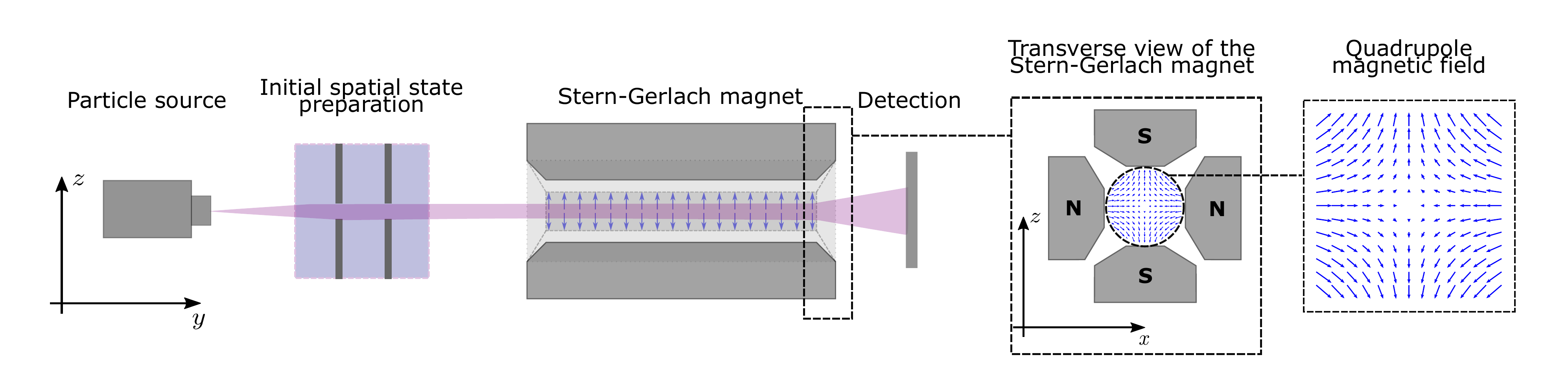}
\caption{ Modified Stern-Gerlach setup for the estimation of the initial spin state of a beam of neutral spin-1/2 particles.
The magnet generates a quadrupolar magnetic field with equal gradients in both the $x$ and $z$ directions.}
\label{fig:sg_setup}
\end{figure*}

The well-known Stern-Gerlach experiment~\cite{Gerlach1922}, in which a beam of neutral particles with definite magnetic dipole moment is made to interact with an external magnetic field, has been widely used to measure the spin projection of the particles along the direction of the field~\cite{Sherwood1954,Hamelin1975,Barkan1968,Jones1980}.
In an approximation in which an inhomogeneous magnetic field with a gradient only in the direction of a large reference field \cite{Bohm1989,Scully1987} is considered, the Stern-Gerlach setup is an ideal spin measurement apparatus~\cite{VonNeumann}.
More complete semiclassical \cite{Cruz2000} and quantum \cite{Potel2005,Hsu2011} descriptions of the experiment, which take into account a second gradient component to satisfy Gauss's law, have shown that the Stern-Gerlach setup is not an ideal spin meter; the magnetic field inhomogeneities cause beam deflections that are not determined only by the spin projections of the particles. 
However, the presence of a strong reference field still allows the correct \textit{estimation} of the initial spin projection along its direction.

Although the purpose of the Stern-Gerlach setup is to measure a spin projection, it is tempting to question if setup modifications can provide more information.
Weigert~\cite{Weigert1992}  showed that projection measurements along two different spatial directions, and one in the direction perpendicular to both of them, enable state reconstruction. 
This method requires changing the direction of the beam without modifying the spin state.
Stern-Gerlach setups lacking the reference magnetic field allow either the estimation of two of the three components of the Bloch vector that defines the initial spin state of the beam~\cite{Muynck2006}, or the projective measurement of a spin component, if the  initial spatial wave function is carefully chosen~\cite{Garraway1999}.
These results can be regarded as a demonstration that a large reference field somewhat limits the information that can be obtained about the spin state of the particles of the beam. 


Could quadrupolar fields allow the estimation of the whole initial spin state?
In a naive view, the quadrupolar field can be seen as two apparatuses which try to measure two orthogonal components of the initial spin state.
In such circumstances, it has been shown~\cite{Saavedra2019} that it is almost always possible to estimate the whole state.
This heuristic analysis points to a positive answer to the previous question.
In fact, as shown in this work (Sec.~\ref{sec:level2}), it is possible to estimate the initial (pure or mixed) spin state of a beam of neutral spin-1/2 particles, using linear inversion or maximum likelihood estimation~\cite{Siah2015}, when the initial spatial state is chosen to be an elongated Gaussian.

Our results are obtained by a combination of numerical and analytical methods.
The time evolution corresponding to the Hamiltonian of the modified Stern-Gerlach setup (Sec.~\ref{sec:level1}) is numerically performed using the Suzuki-Trotter method  (App.~\ref{sec:app3}).
The error of the proposed state estimation procedures is quantified by the logarithmic error $\Delta(\mathcal{G},\boldsymbol{s})$ of the scheme, defined in Sec.~\ref{sec:level4}.
Although the logarithmic error greatly varies from one set of parameters to another, and also depends on the initial state, it is reasonably low in some regions of the parameter space which are within reach of current experimental techniques (Sec.~\ref{sec:conc}).

\section{\label{sec:level1}Description of the model}

Consider the Stern-Gerlach setup shown in Fig.~\ref{fig:sg_setup}: A beam of neutral spin-1/2 particles of mass $m$ and magnetic dipole moment $\mu$ is prepared in a particular factorized initial state, $\rho(0) = R(0)\rho_S(0),$ where $R$ describes the spatial state and $\rho_S$ the spin state of the particles.
The spatial state is considered to be completely defined, while the spin state is taken to be unknown.
The particles of the beam are identical, indistinguishable, independent and far enough apart from each other, that any interaction between them can be ignored. 
After preparation, the particles are sent through an inhomogeneous magnetic field $\vec{B}$, generated by a magnet of length $L$, which deflects the beam.
The magnetic field is assumed to have components only on the plane $(x,z)$, perpendicular to the propagation direction of the beam. 
Border effects are ignored.
After the interaction with the magnetic field, the beam might evolve freely for some time before finally being detected on a screen. 

The Hamiltonian for each particle of the beam in the presence of the magnetic field is  
\begin{equation}
{H}=\frac{p_y^2}{2m}
+\left(\frac{p_x^2+p_z^2}{2m}
+\mu\vec{S}\cdot\vec{B}(x,z)\right)
=H_{y} + H_{xz},
\label{eq:hamiltonian}
\end{equation}
where the subscripts indicate the dependence on the spatial coordinates. 
The corresponding time evolution operator $U(t,0)=\exp(-iH t/\hbar)$, can be written as $U(t,0) = U_{xz}(t,0)  U_{y}(t,0)$, where $U_{xz}(t,0)$ and $U_{y}(t,0)$ are the time evolution operators corresponding to $H_{xz}$ and $H_{y}$, respectively. 

A factorized initial spatial state of the particles, $R(0) = R_{xz}(0)   R_{y}(0)$, allows the separation of the dynamics along $y$, the longitudinal coordinate. 
Since this dynamics corresponds to a free evolution, the time spent on the magnetic field region, $\tau$, can be approximated by $\tau = mL/\hbar k_{0y}$, where it has been assumed that the momentum distribution in the $y$ coordinate is strongly peaked around $\hbar k_{0y}.$
Our study reduces to a two-dimensional problem because further influence of the time evolution in the longitudinal coordinate can be ignored. 
The remaining part of the initial spatial state, $R_{xz}(0)$, is assumed to be of the form $R_{xz}(0) = \ket{\psi_{xz}} \bra{\psi_{xz}}$ where
$$\braket{x,z|\psi_{xz}} = 
\sqrt{\frac{1}{2\pi\sigma \sigma^\prime}}
\exp\left[-\left(\frac{x^2}{4\sigma^2}+\frac{z^2}{4\sigma'^2}\right)\right].$$

The transit time, $\tau$, and the dispersion of the initial spatial state in the $x$-direction, $\sigma$, are used as natural scales to define the dimensionless quantities  $\bar{t}=t/\tau$, $\bar{x}=x/\sigma$, $\bar{z}=z/\sigma$, $\bar{p}_x=\sigma p_x/\hbar$, $\bar{p}_z=\sigma p_z/\hbar$ and $\bar{H}_{\bar{x}\bar{z}} =\tau H_{xz} /\hbar$.  

With these definitions, the equation of motion for the time evolution operator $U_{\bar{x}\bar{z}}(\bar{t},0)$ becomes 
$$
    i\frac{dU_{\bar{x}\bar{z}} (\bar{t},0)}{d\bar{t}} = \bar{H}_{\bar{x}\bar{z}}  U_{\bar{x}\bar{z}} (\bar{t},0),
$$
where
\begin{equation}
   \bar{H}_{\bar{x}\bar{z}} =g_2\left(\bar{p}^2_x+\bar{p}^2_z\right)+g_1\left(\bar{x}\sigma_1-\bar{z}\sigma_3\right),
\label{eq:dimesionless_hamiltonian}   
\end{equation}
$g_1=\mu b\sigma\tau/2\hbar$, $g_2=\hbar\tau/2m\sigma^2$ and $\{\sigma_i\}$, $i=1,2,3$, stand for the Pauli spin operators. 
For this model, we have considered a quadrupolar magnetic field of the form
\begin{equation}
\vec{B}(\bar{x},\bar{z})=- b\sigma\left(\bar{x}\hat{\imath}-\bar{z}\hat{k}\right).
\label{eq:magnetic_field}
\end{equation}
In the scaled coordinates, the initial Gaussian wavefunction reads
\begin{equation}
\braket{\bar{x},\bar{z}|\bar{\psi}_{\bar{x}\bar{z}}} = 
\sqrt{\frac{1}{2\pi\lambda}}
\exp\left[-\frac{\bar{x}^2+\left(\bar{z}/\lambda\right)^2}{4} \right],
\label{eq:dimesionless_initial_state}
\end{equation}
where $\lambda=\sigma'/\sigma$.

The free evolution of the beam just after the interaction with the magnetic field and before being detected at time $T$ is represented by the operator $U_{\bar{x}\bar{z}}^{(f)}(T,1)=\exp[-i\bar{H}_{\bar{x}\bar{z}}^{(f)}(T-1)],$ where $\bar{H}_{\bar{x}\bar{z}}^{(f)}=g_2\left(\bar{p}^2_x+\bar{p}^2_z\right)$. 
By combining the free and magnetic parts of the time evolution, we obtain the time evolution operator for the complete Stern-Gerlach setup, $U_{\bar{x}\bar{z}}^{(t)}(T,0) = U_{\bar{x}\bar{z}}^{(f)}(T,1)U_{\bar{x}\bar{z}} (1,0)$. 
The final state of the particles of the beam just before detection will then be
\begin{equation}
\rho(T)=U_{\bar{x}\bar{z}}^{(t)}(T,0)\left(R_{xz}(0) \rho_S(0)\right)U_{\bar{x}\bar{z}}^ {(t)\dagger}(T,0).    
\label{eq:final_state}
\end{equation}
From here on, we will drop the bar on top of the dimensionless variables to unclutter the notation. 
The Stern-Gerlach setup described in this work can be characterized by a set $\mathcal{G}=\{ g_1, g_2, \lambda, T\}$ of dimensionless parameters: $g_1$ and $g_2$, associated with the quadrupole field and the kinetic energy, respectively; $\lambda$, which measures the elongation of the initial spatial wavefunction; and $T-1$, the time interval of free-evolution after interaction with the magnetic field.

In the usual theoretical treatments of the Stern-Gerlach experiment, it is assumed that the inhomogeneous magnetic field has a large constant reference component; for example, in the $z$ direction.
This field component allows to neglect the term $g_1x\sigma_1$ in the corresponding Hamiltonian, for sufficiently localized spatial states in the neighborhood of $x=0$.
Under this approximation, the complete time evolution operator commutes with $\sigma_3$, and  the $z$-component of the spin of the particles~\cite{Muynck2006} can be measured using the spatial degrees of freedom, which have become correlated with this spin degree of freedom ($\sigma_3$, in this case).

Correlations are established between spatial and spin degrees of freedom, even in the absence of the reference magnetic field.
Hence, in our model, the spatial intensity distribution on the screen $I(x,z) = \operatorname{Tr}_S\left(\braket{x,z|\rho(T)|x,z}\right)$, would still contain information about the initial spin state of the beam.
Here, $\operatorname{Tr}_S\left(\cdot\right)$ denotes the trace over the spin degrees of freedom.

In the next section, we show that information about the complete initial spin state of the beam is indeed contained in the spatial intensity distribution of the beam. We then investigate how to use this intensity measurements to estimate the initial spin state of the particles.

\section{\label{sec:level2}State estimation}

It is intuitively reasonable that spin state estimation using measurements of the spatial intensity distribution should be possible under fairly general conditions. 
However, as detailed in Appendix~\ref{sec:app1}, initial spatial states which remain invariant under rotations around the propagation direction of the beam ($\lambda=1$ in Eq.~\eqref{eq:dimesionless_initial_state}) do not encode information about the second spin component.
Consequently,  the complete estimation of the initial spin state of the beam from measurements of its spatial intensity distribution requires us to assume that $\lambda$ is different from unity.

For a general final state $\rho(T)$, the intensity at a point $(x,z)$ on the screen can be written as
\begin{eqnarray}
    I(x,z) =  \operatorname{Tr} \left[ \rho(T)\ket{x,z}\bra{x,z}\sigma_0 \right] = \operatorname{Tr} \left[\rho(T)Q_{xz}\right]  ,
    \label{eq:intensity_at_point}
\end{eqnarray}
where $Q_{xz}$ was defined as $Q_{xz} = \ket{x,z}\bra{x,z}\sigma_0.$
Using Eq.~\eqref{eq:final_state}, we can express this distribution in the form
\begin{align}
\begin{split}
I(x,z) &= \textrm{Tr}\left[U_{xz}^{(t)}(T,0)R_{xz}(0)\rho_S(0)U_{xz}^{(t)\dagger}(T,0)Q_{xz}\right]\\
&= \textrm{Tr}\left[U^{(t)\dagger}_{xz}(T,0)Q_{xz}U_{xz}^{(t)}(T,0)R_{xz}(0)\rho_S(0)\right]\\
&=\textrm{Tr}_S[\tilde{Q}_{xz}(T)\rho_S(0)],
\end{split}
\label{eq:intensity_as_spin_measurement}
\end{align}
where $\tilde{Q}_{xz}(T)=\textrm{Tr}_{xz}\left[U^{(t)\dagger}_{xz}(T,0)Q_{xz}U_{xz}^{(t)}(T,0)R_{xz}(0)\right]$ and $\textrm{Tr}_{xz}(\cdot)$ indicates the trace over the spatial degrees of freedom. 
In this way, we can interpret intensity at point $(x,z)$ as a measurement of the spin observable $\tilde{Q}_{xz}(T)$ over the initial spin state of the beam. 
The complete set of operators $\tilde{Q}_{xz}(T)$ represents the whole Stern-Gerlach setup: spatial preparation, time evolution and intensity measurement.

Both $\rho_S(0)$ and  $\tilde{Q}_{xz}(T)$ can be expanded in the basis $\{\sigma_\mu\}$, 
\begin{eqnarray}
\rho_S(0)&=&\frac{1}{2}\sum_{\mu=0}^3s_\mu \sigma_\mu,
\label{eq:initial_spin_state_pauli}\\
\tilde{Q}_{xz}(T) &=&\sum_{\mu=0}^3 M_{\mu}(x,z,T)\sigma_\mu.
\label{eq:int_meas_op_pauli}
\end{eqnarray}
Using these expressions into Eq.~\eqref{eq:intensity_as_spin_measurement}, we see that
\begin{equation}
I(x,z)=\sum_{\mu=0}^3M_\mu(x,z,T)s_\mu,
\label{eq:intenisty_bloch}    
\end{equation}
relates the intensity measurements with the real parameters $s_\mu,$ where $s_0=\textrm{Tr}_S[\rho_S(0)]$ and the Bloch vector components $s_\mu$, $\mu=1,2,3$, define the initial spin state.
Since $\tilde{Q}_{xz}(T)$ is Hermitian, coefficients $M_{\mu}(x,z,T)$ are also real.
 

\begin{figure}[htbp!]
\centering
\includegraphics[width=0.8\columnwidth]{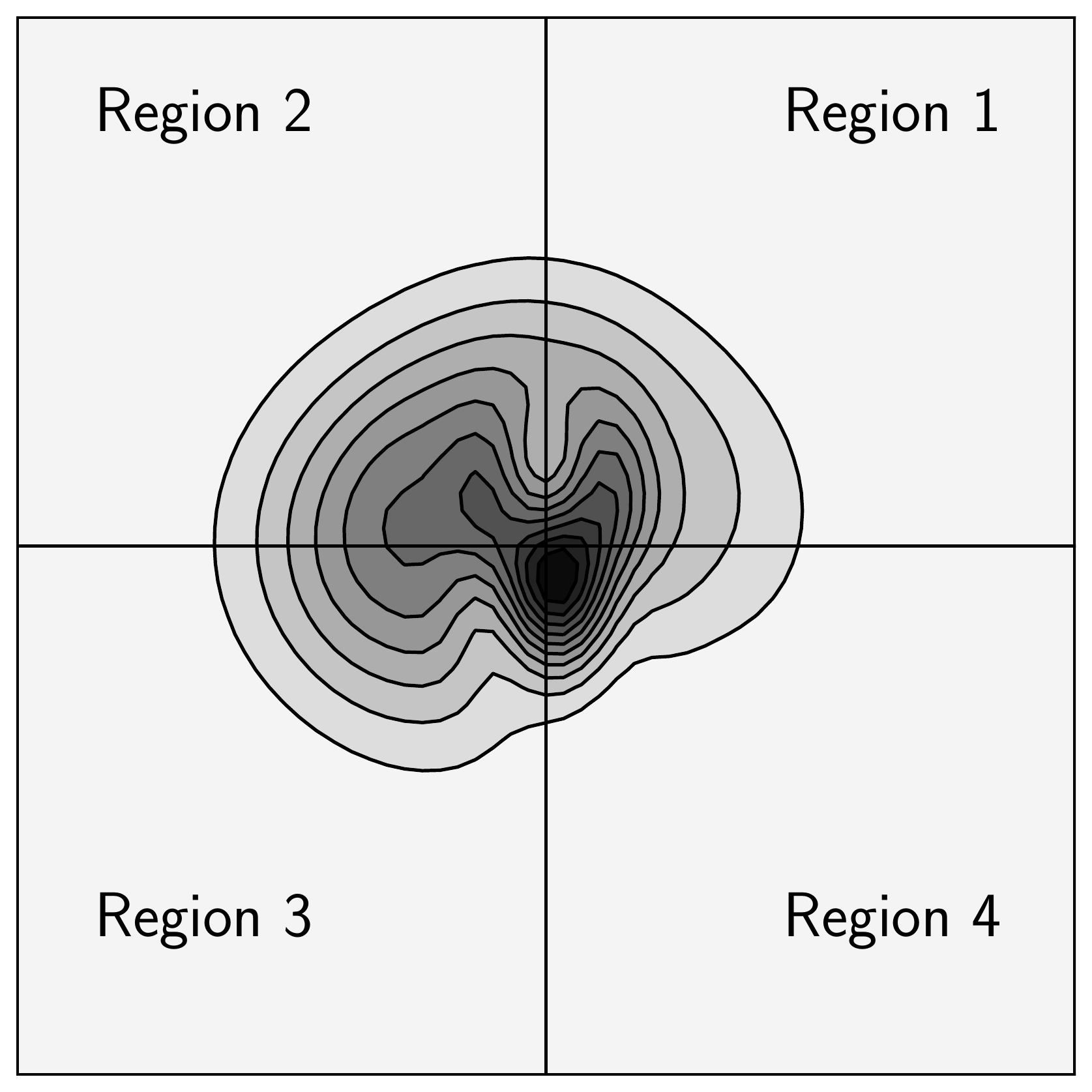}
\caption{Chosen regions over the $(x,z)$ plane for the definition of the intensity measurement operators necessary for the estimation of the initial spin state of the beam. The division of the detection screen is similar to the design of quadrant detectors.}
\label{fig:sg_regions}
\end{figure}

We will consider two different approaches for the estimation of the initial spin state. 
The first one, in the spirit of the original Stern-Gerlach setup, divides the intensity measurement in quadrants. 
This measurement will be represented by four observables, the minimum number of operators required for the complete estimation of the state~\cite{Siah2015}; since the spin state can be defined by three real parameters, at least four different measurements are necessary for their estimation. 
The second approach uses the intensity at every point on the screen to reconstruct the initial spin state.

\subsection{\label{sec:quad_app}Quadrant approach}

We arbitrarily choose the four regions $\Omega_k,$ $k=1,2,3,4,$ shown in Fig.~\ref{fig:sg_regions} for the measurement of the spatial intensity distribution of the beam. 
Each one of these measurements will be represented by a spin operator 
\begin{equation}
\tilde{Q}_k(T)=\int_{\Omega_k}\tilde{Q}_{xz}(T)\,d{x}d{z}.
\label{eq:measurement_operators_S}
\end{equation}
Since the set $\{\tilde{Q}_k(T)\}$ constitutes a POVM (Positive Operator-Valued Measure), the intensities $p_k(T)=\operatorname{Tr}_S[\rho_S(0)\tilde{Q}_k(T)]$ can be interpreted as the probabilities of detection of one particle at each region. 


Following a similar procedure to the one leading to Eq.~\eqref{eq:intenisty_bloch}, we can express the theoretical probabilities of detection as
\begin{equation}
p_k(T)=\sum_{\mu=0}^3M_{k\mu}(T)s_\mu,
\label{eq:probability_dist_param} 
\end{equation}
where
\begin{equation}
M_{k\mu}(T)=\int_{\Omega_k}M_\mu(x,z,T)\,dxdz
\label{eq:mesaurement_matrix_components}    
\end{equation}
are the coefficients of the expansion of operators $\tilde{Q}_k(T)$ in terms of Pauli spin operators. Defining the vectors $\boldsymbol{p}=(p_1,p_2,p_3,p_4)$ and $\boldsymbol{s}=(s_0,s_1,s_2,s_3)$, Eq.~\eqref{eq:probability_dist_param} takes the  simple matrix form
\begin{equation}
\boldsymbol{p}(T)=\mathbb{M}(T)\boldsymbol{s},
\label{eq:pd_matrix_form}
\end{equation}
where $\mathbb{M}(T)$, whose elements are $M_{k\mu}(T)$, is called the \textit{measurement matrix} of the system.

To estimate the components of the Bloch vector, we must relate probabilities $p_k(T)$, and therefore parameters $s_\mu$, to the outcomes of the measuring process. If we consider the beam to be formed by $N$ particles, these outcomes correspond to the number of particles $n_k$ detected at region $k$, where $\sum_{k=1}^4n_k=N$. The values $\boldsymbol{n}=\{n_1,n_2,n_3,n_4\}$ constitute a set of random variables whose joint probability distribution is a multinomial distribution of the form~\cite{Watanabe2013} 
\begin{equation}
P(\boldsymbol{n}|\boldsymbol{s})=N!\prod_{k=1}^4\frac{1}{n_k!}\left[p_k(T)\right]^{n_k}.
\label{eq:multinomial_dist}    
\end{equation}
The dependence of this distribution with respect to $\boldsymbol{s}$ comes from relation~\eqref{eq:probability_dist_param}. 

When seen as a function of the unknown state $\rho_S(0)$, instead of a function of $\boldsymbol{n}$, distribution~\eqref{eq:multinomial_dist} corresponds to the likelihood function of the state. By maximizing this function with respect to $\rho_S(0)$ and under the adequate constraints, we can find a maximum-likelihood estimator for the components $s_\mu$~\cite{Siah2015,Paris2004}.

It is usually simpler to maximize the natural logarithm of the likelihood function, i.e. the log-likelihood function. A detailed exposition of the maximization of this function is presented in Appendix~\ref{sec:app2}.

By considering only that the estimated state should be normalized, we obtain the following relation between the observed frequencies $f_k=n_k/N$ and $\check{s}_\mu,$ the estimators of $s_\mu$:
\begin{equation}
f_k=\sum_{\mu=0}^3M_{k\mu}\check{s}_\mu.
\label{eq:linear_inversion_estimator}    
\end{equation}
As in the case of equation~\eqref{eq:pd_matrix_form}, relation~\eqref{eq:linear_inversion_estimator} can be written as $\boldsymbol{f}=\mathbb{M}\boldsymbol{\check{s}}$, where $\boldsymbol{f}=(f_1,f_2,f_3,f_4)$ and $\boldsymbol{\check{s}}=(\check{s}_0,\check{s}_1,\check{s}_2,\check{s}_3)$. Consequently, $\boldsymbol{\check{s}}$ is easily calculated as 
\begin{equation}
\boldsymbol{\check{s}}=\mathbb{M}^{-1}(T)\boldsymbol{f}.
\label{eq:linear_inversion_estimator_mx}    
\end{equation}

This estimator corresponds to a \textit{linear inversion estimator}~\cite{Siah2015}. In a more general setup, when the intensity distribution is measured in more than four regions and the measurement matrix becomes non-square, Eq.~\eqref{eq:linear_inversion_estimator_mx} holds if $\mathbb{M}^{-1}(T)$ is interpreted as the Moore-Penrose inverse. 

Although estimator~\eqref{eq:linear_inversion_estimator_mx} has a simple analytical expression in terms of the measurement results, it could lead to non-physical estimations of the initial spin state; a real state should not only be normalized but also positive semidefinite. Considering this constraint, the estimators for the initial spin state satisfy $\check{s}_0=1$ and
\begin{equation}
(1-\check{r}_0)\check{s}_\mu=\check{r}_\mu
\label{eq:ML_estimator_th}    
\end{equation}
for $\mu=1,2,3$, where
\begin{equation}
\check{r}_\mu=\sum_{k=1}^4\frac{f_k}{\check{p}_k(T)}M_{k\mu}(T)
\label{eq:ML_estimator_coeff}    
\end{equation}
and values $\check{p}_k(T)$ are calculated from Eq.~\eqref{eq:probability_dist_param} using estimators $\check{s}_\mu$ instead of parameters $s_\mu$.

Relation~\eqref{eq:ML_estimator_th} is highly non-linear and cannot, in general, be solved by analytical means. However, there are several algorithms for its numerical computation~\cite{Paris2004,Rehaceck2007}. Here we will use the ``$R\rho R$ algorithm''~\cite{Rehaceck2007}, which allows the iterative computation of the estimators. The details of the derivation of this method are presented in Appendix~\ref{sec:app2}. The estimators of the initial spin state, for $\mu=1,2,3$, are calculated as follows:
\begin{equation}
\check{s}_\mu^{(n+1)}=\frac{2\check{r}_\mu^{(n)} - \check{s}_\mu^{(n)}\check{\gamma}^{(n)}}{2\check{r}_0^{(n)} + \check{\gamma}^{(n)}},
\label{eq:iterative_MLE}
\end{equation}
where the superscript $(n)$ indicates the iteration step and $$\check{\gamma}^{(n)}=\sum_{\mu=1}^3\left(\check{r}_\mu^{(n)}\right)^2-\left(\check{r}_0^{(n)}\right)^2.$$ Each iteration results in a normalized estimated state, so that $\check{s}_0^{(n)}=1$ for every $n$. 

The numerical implementation of this algorithm requires the previous definition of the initial estimated state. This state is usually taken to be the maximally mixed spin state defined by  $\check{s}_0^{(0)}=1$ and $\check{s}_\mu^{(0)}=0$ for $\mu=1,2,3$.

\subsection{\label{sec:wd_app}Continuous distribution approach}
When we consider the complete intensity distribution over the screen, the outcomes of the measuring process are not interpreted as a set of particle detections in a given region. Instead, the results of the measurement will correspond to a set of coordinates $\boldsymbol{v}=\{(x_k,z_k)\}$, where each pair indicates that a particle is detected in a small region around the position $(x_k,z_k)$ on the screen.

The set $\boldsymbol{v}$ constitutes a set of two-dimensional, continuous random variables whose joint probability density function is given by 
\begin{equation}
F\left(\boldsymbol{v}|\boldsymbol{s}\right)=\prod_{k=1}^NI(x_k,z_k),
\label{eq:cont_density_function}    
\end{equation}
where, as in relation~\eqref{eq:intenisty_bloch}, $I(x,z)=\sum_{\mu=0}^3M_\mu(x,z,T)s_\mu$.

When seen as a function of $\rho_S(0)$ instead of variables $(x_k,z_k)$, $F\left(\boldsymbol{v}|\boldsymbol{s}\right)$ is interpreted as the likelihood function of the state. By maximizing the corresponding log-likelihood function under the considerations that the estimated state should be normalized and positive semidefinite (see Appendix~\ref{sec:app2}), the estimators for the initial spin state fulfill the relations $\check{s}_0=1$ and
\begin{equation}
(1-\check{R}_0)\check{s}_\mu=\check{R}_\mu
\label{eq:ML_estimator_cont}    
\end{equation}
for $\mu=1,2,3$, where
\begin{equation}
\check{R}_\mu=\frac{1}{N}\sum_{k=1}^N\frac{M_\mu(x_k,z_k,T)}{\check{I}(x_k,z_k)}.
\label{eq:ML_estimator_cont_coeff}    
\end{equation}
The function $\check{I}(x,z)$ is calculated from relation~\eqref{eq:intenisty_bloch} using estimators $\check{s}_\mu$ instead of parameters $s_\mu$.

As in the case of the quadrant approach, equation~\eqref{eq:ML_estimator_cont} cannot, in general, be solved analytically. By using the $R\rho R$ algorithm, the estimators $\check{s}_\mu$, for $\mu=1,2,3$, can be calculated numerically from the iterative relation
\begin{equation}
\check{s}_\mu^{(n+1)}=\frac{2\check{R}_\mu^{(n)} - \check{s}_\mu^{(n)}\check{\Gamma}^{(n)}}{2\check{R}_0^{(n)} + \check{\Gamma}^{(n)}},
\label{eq:iterative_MLE_cont}
\end{equation}
where $$\check{\Gamma}^{(n)}=\sum_{\mu=1}^3\left(\check{R}_\mu^{(n)}\right)^2-\left(\check{R}_0^{(n)}\right)^2.$$ As before, each iteration results in a normalized estimated state ($\check{s}_0^{(n)}=1$) and the computation begins by choosing a maximally mixed initial state.

In the remainder of the document, we will refer to estimator~\eqref{eq:linear_inversion_estimator_mx} as the \textit{linear inversion estimator} of the initial spin state, while estimators~\eqref{eq:ML_estimator_th} and~\eqref{eq:ML_estimator_cont}, and their corresponding numerical versions Eqs.~\eqref{eq:iterative_MLE} and~\eqref{eq:iterative_MLE_cont}, will be referred to as the \textit{discrete and continuous maximum-likelihood estimators} of the initial spin state, respectively.

\section{\label{sec:level4}Error of the estimation}
In principle, the linear inversion and maximum-likelihood estimators, allow the estimation of all the parameters defining the initial spin state. However, it is necessary to evaluate how reliable the estimation of these parameters can actually be. 
It is expected, for example, that the estimation of $s_2$ becomes increasingly difficult as $\lambda$ approaches unity. 
A suitable state estimation thus requires a proper choice of the setup parameters $\mathcal{G} =\{g_1, g_2, \lambda, T\}$. 
To investigate this problem, we will quantify the error of the estimation and analyze its dependence on the setup parameters. 

Statistically, the performance of estimators $\check{s}_\mu$ is characterized by their bias, $b(\check{s}_\mu)=\mathbb{E}[\check{s}_\mu-s_\mu]$, and their covariance matrix $\textrm{Cov}(\boldsymbol{\check{s}},\boldsymbol{\check{s}})$~\cite{Kay1993}, whose elements are defined by $$\textrm{Cov}(\check{s}_\mu,\check{s}_\nu)=\mathbb{E}\left[\check{s}_\mu\check{s}_\nu\right]-\mathbb{E}\left[\check{s}_\mu\right]\mathbb{E}\left[\check{s}_\nu\right].$$ The symbol $\mathbb{E}[\cdot]$ indicates a statistical expectation value with respect the corresponding probability distribution or probability density function. The diagonal elements of the covariance matrix, i.e. the variances, are associated with the error of the estimated parameters, $s_\mu = \mathbb{E}[\check{s}_\mu] \pm \sqrt{\operatorname{Cov}(\check{s}_\mu,\check{s}_\mu)}$. Good estimators for the initial spin state should have small values of $|b(\check{s}_\mu)|$ and $\textrm{Cov}(\check{s}_\mu,\check{s}_\mu)$.

The linear inversion estimator is unbiased, $b(\check{s}_\mu)=0$ for all $\mu$, so its performance is determined entirely by its covariance matrix. The maximum-likelihood estimators, on the other hand, are asymptotically unbiased, thus, by considering a large enough number of particles, their performance will also be determined their respective covariance matrices~\cite{Kay1993}.

The maximum performance of the state estimation schemes, corresponding to the minimum values that the variances can take, will be characterized by the inverse of their \emph{Fisher information matrix} (or simply \textit{information matrix}), $\mathbb{J}\left(\boldsymbol{s}\right)$.
For the linear inversion estimator, this characterization is justified because any unbiased estimator of $\boldsymbol{s}$ satisfies the Cramér-Rao lower bound~\cite{Kay1993}
\begin{equation}
\textrm{Cov}\left(\boldsymbol{\check{s}},\boldsymbol{\check{s}}\right)-\mathbb{J}^{-1}\left(\boldsymbol{s}\right)\geq 0,
\label{eq.Cramer-Rao_ineq}    
\end{equation}
where the inequality indicates that the difference between matrices is positive semidefinite. Therefore, $\mathbb{J}^{-1}(\boldsymbol{s})$ is the lowest possible covariance matrix associated to the linear inversion estimation procedure.

For the maximum-likelihood estimators, the use of the information matrix is justified by noticing that they are asymptotically efficient, thus, for a large enough number of particles, their respective covariance matrices will correspond to $\mathbb{J}^{-1}(\boldsymbol{s})$~\cite{Kay1993}:
\begin{equation}
\lim_{N\rightarrow\infty}\textrm{Cov}(\boldsymbol{\check{s}},\boldsymbol{\check{s}})=\mathbb{J}^{-1}(\boldsymbol{s}).
\label{eq:asymptotic_efficiency}
\end{equation}

The information matrix does not depend on the construction of the estimator of the spin state, it only depends on the corresponding probability distribution or probability density function. As a consequence, the linear inversion and discrete maximum-likelihood estimators, i.e. the quadrant approach estimation procedures, will have the same information matrix, whose elements are calculated as 
\begin{equation}
J_{\mu\nu}(\boldsymbol{s})=-\mathbb{E}\left[\frac{\partial^2\ln P(\boldsymbol{n}|\boldsymbol{s})}{\partial s_\mu\partial s_\nu}\right],
\label{eq:fisher_info_disc}
\end{equation}
where $P(\boldsymbol{n}|\boldsymbol{s})$ is the multinomial distribution given by Eq.~\eqref{eq:multinomial_dist}. 
A direct computation of these elements show that they can be written in terms of the elements of the measurement matrix, $M_{k\mu}(T)$, as
\begin{equation}
J_{\mu\nu}(\boldsymbol{s})=N \sum_{k=1}^4\frac{1}{p_k}M_{k\mu}(T)M_{k\nu}(T)=NK_{\mu\nu}(\boldsymbol{s})\label{eq:fisher_components_disc},
\end{equation}
where the dependence of elements $K_{\mu\nu}(\boldsymbol{s})$ on $\boldsymbol{s}$ comes form relation $p_k(T)=\sum_{\mu=0}^3M_{k\mu}(T)s_\mu$. 

For the continuous maximum-likelihood estimator, the components of the information matrix are given by
\begin{equation}
J_{\mu\nu}(\boldsymbol{s})=-\mathbb{E}\left[\frac{\partial^2\ln F(\boldsymbol{v}|\boldsymbol{s})}{\partial s_\mu\partial s_\nu}\right],
\label{eq:fisher_info_cont}
\end{equation}
where $F(\boldsymbol{v}|\boldsymbol{s})$ is the probability density function given by Eq.~\eqref{eq:cont_density_function}. As in the case of the linear inversion and discrete maximum-likelihood estimators, a direct calculation of the elements $J_{\mu\nu}(\boldsymbol{s})$ reveals that they can be written in terms of functions $M_\mu(x,z,T)$ as
\begin{align}
J_{\mu\nu}(\boldsymbol{s})&=N\int_{-\infty}^{\infty}\frac{1}{I(x,z)}M_\mu(x,z,T)M_\nu(x,z,T)\,dxdz\nonumber\\
&=NK_{\mu\nu}(\boldsymbol{s})\label{eq:fisher_components_cont}, 
\end{align}
where, in this case, the dependence of elements $K_{\mu\nu}(\boldsymbol{s})$ on $\boldsymbol{s}$ comes form relation $I(x,z)=\sum_{\mu=0}^3M_\mu(x,z,T)s_\mu$. 

To eliminate the dependence of the information matrix on the number of runs of the experiment, we consider the scaled information matrix $\mathbb{K}(\boldsymbol{s}) = \mathbb{J}(\boldsymbol{s})/N$, whose elements are the values $K_{\mu\nu}(\boldsymbol{s})$ introduced in Eqs.~\eqref{eq:fisher_components_disc} and~\eqref{eq:fisher_components_cont}.
Since $\mathbb{J}^{-1}(\boldsymbol{s})$ decreases at a rate $N^{-1}$, we can achieve a desired value for the variances by choosing a large but adequate number of particles. 
However, the choice of $N$ will be strongly limited by how large the diagonal elements of $\mathbb{K}^{-1}(\boldsymbol{s})$ are. 
For this reason, we will ignore the explicit presence of the number of particles and define the error of the estimation procedure as a function of these diagonal elements.

\begin{figure*}[hbtp!]
\centering
\includegraphics[width=0.89\linewidth]{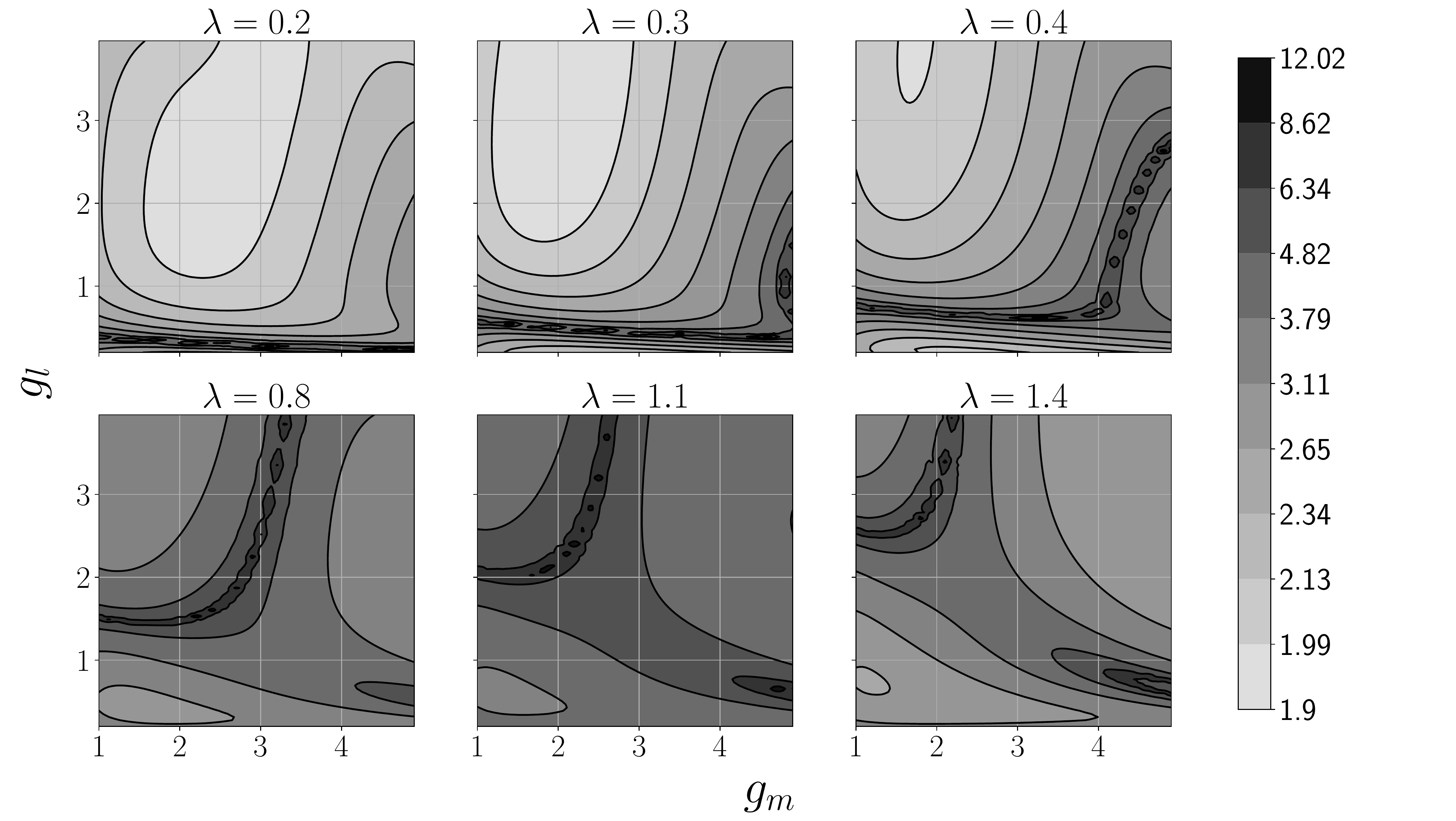}
\caption{Error of the linear inversion and discrete maximum-likelihood estimation procedures, $\Delta(\mathcal{G},\boldsymbol{s})$, as function of parameters $g_1$ and $g_2$ for different values of parameter $\lambda$. 
The beam was assumed to be detected just at the end of the interaction with the magnetic field, that is at $T=1.0$.
The initial spin state of the beam was a pure state defined by its Bloch vector $\boldsymbol{s}=(\sin\theta\cos\phi, \sin\theta\sin\phi, \cos\theta)$, where $\theta=1.91$ and $\phi=4.78$.}
\label{fig:trace_distance_th}
\end{figure*}

\begin{figure*}[hbtp!]
\centering
\includegraphics[width=0.89\linewidth]{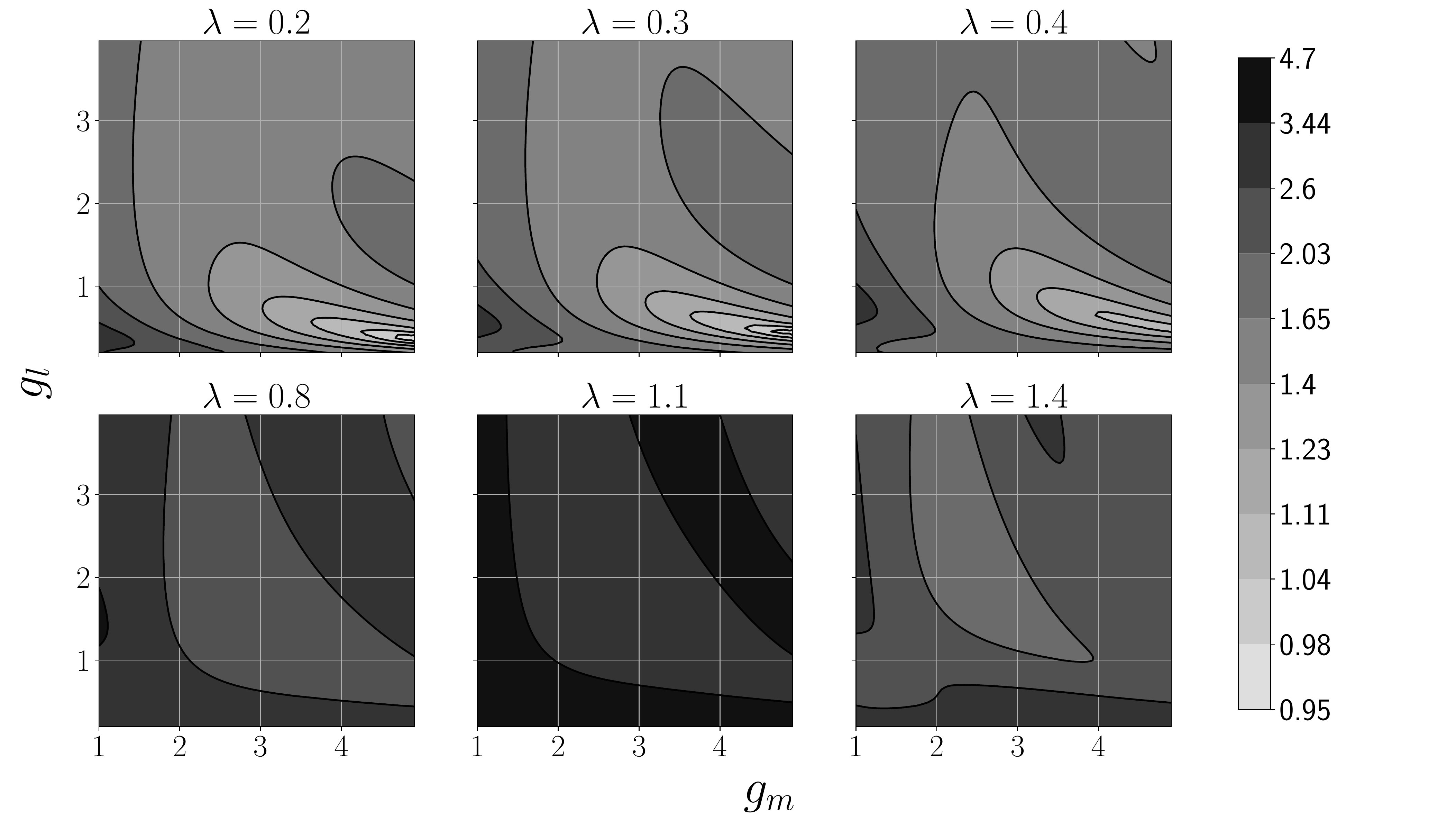}
\caption{Error of the continuous maximum-likelihood estimation procedure, $\Delta(\mathcal{G},\boldsymbol{s})$, as function of parameters $g_1$ and $g_2$ for different values of parameter $\lambda$. 
The beam was assumed to be detected just at the end of the interaction with the magnetic field, that is at $T=1.0$.
The initial spin state of the beam was a pure state defined by its Bloch vector $\boldsymbol{s}=(\sin\theta\cos\phi, \sin\theta\sin\phi, \cos\theta)$, where $\theta=1.91$ and $\phi=4.78$.}
\label{fig:trace_distance_th_cont}
\end{figure*}

\begin{figure*}[hbtp!]
\centering
\includegraphics[width=0.89\linewidth]{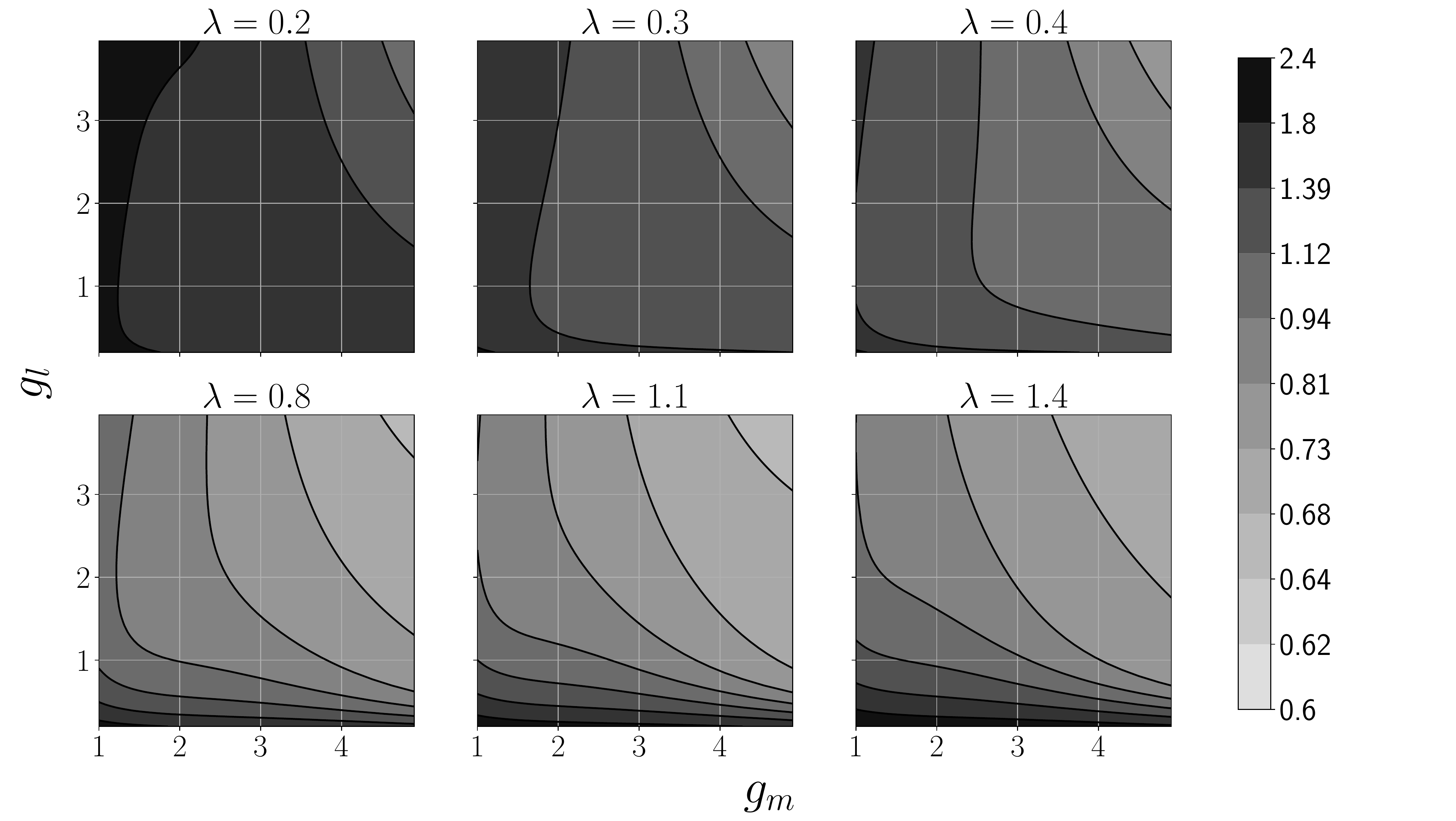}
\caption{Error of the linear inversion and discrete maximum-likelihood estimation procedures, $\Delta(\mathcal{G},\boldsymbol{s})$, as function of parameters $g_1$ and $g_2$ for different values of parameter $\lambda$, without including the variance corresponding to the estimation of $s_2$. 
The beam was assumed to be detected just at the end of the interaction with the magnetic field, that is at $T=1.0$. 
The initial spin state of the beam was a pure state defined by the values $\theta=1.91$ and $\phi=4.78$.}
\label{fig:trace_distance_tw}
\end{figure*}

\begin{figure*}[hbtp!]
\centering
\includegraphics[width=0.89\linewidth]{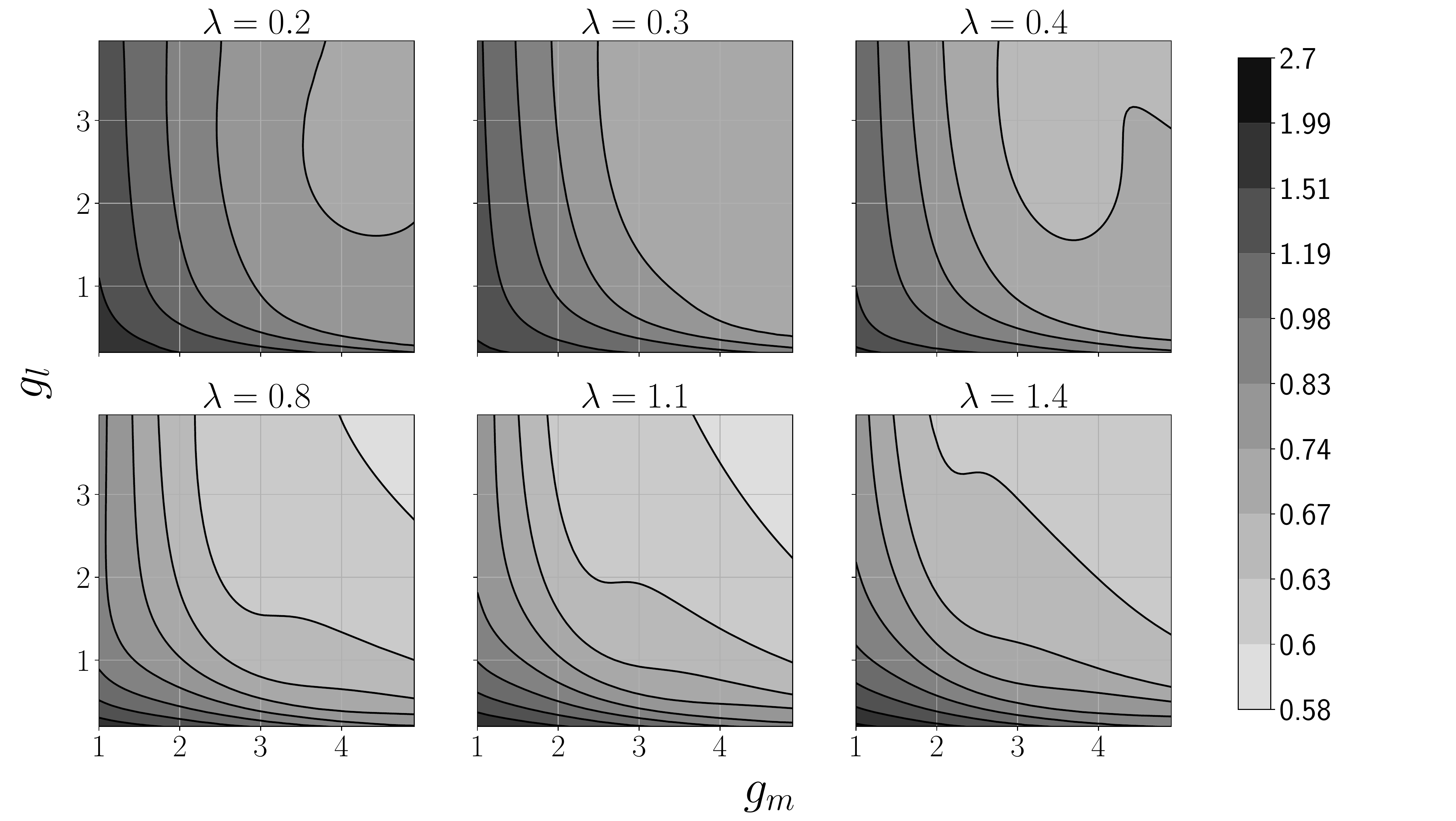}
\caption{Error of the continuous maximum-likelihood estimation procedure, $\Delta(\mathcal{G},\boldsymbol{s})$, as function of parameters $g_1$ and $g_2$ for different values of parameter $\lambda$, without including the variance corresponding to the estimation of $s_2$. 
The beam was assumed to be detected just at the end of the interaction with the magnetic field, that is at $T=1.0$. 
The initial spin state of the beam was a pure state defined by the values $\theta=1.91$ and $\phi=4.78$.}\label{fig:trace_distance_tw_cont}
\end{figure*}

To quantify the quality of the estimation, we define the logarithmic error  
\begin{equation}
\Delta(\mathcal{G},\boldsymbol{s})=\textrm{log}_{10}\left[\textrm{tr}\left(\mathbb{K}^{-1}(\boldsymbol{s})\right)\right],
\label{eq:quality_measure}    
\end{equation}
where we use the symbol $\textrm{tr}(\cdot)$ to distinguish the trace of matrix from the trace of an operator, indicated by $\textrm{Tr}(\cdot)$.
The logarithmic error depends not only on the state parameters $\boldsymbol{s}$, but also on the set of dimensionless parameters $\mathcal{G}=\{g_1,g_2,\lambda,T\}$ of the experimental setup.   
This dependence comes from the elements $M_{k\mu}(T)$ and the functions $M_\mu(x,z,T)$, but also from probabilities $p_k(T)$ and the intensity distribution $I(x,z)$.
The logarithmic scale is useful for large variances, like those that are expected for values of $\lambda$ around unity.

To study the performance of the linear inversion and maximum-likelihood estimators, we will assume that the initial spin state is normalized, $s_0=\textrm{Tr}_S[\rho_S(0)]=1$. As a result, the partial derivatives with respect to $s_0$ in the definition of elements $J_{\mu\nu}(\boldsymbol{s})$ are not taken into account, so we can ignore the elements with $\mu=0$ and $\nu=0$ in Eqs.~\eqref{eq:fisher_components_disc} and~\eqref{eq:fisher_components_cont} and $\mathbb{K}(\boldsymbol{s})$ becomes a $3\times3$ matrix in each case.

\begin{figure*}[htbp!]
\centering
\subfloat{
\includegraphics[width=1.0\columnwidth]{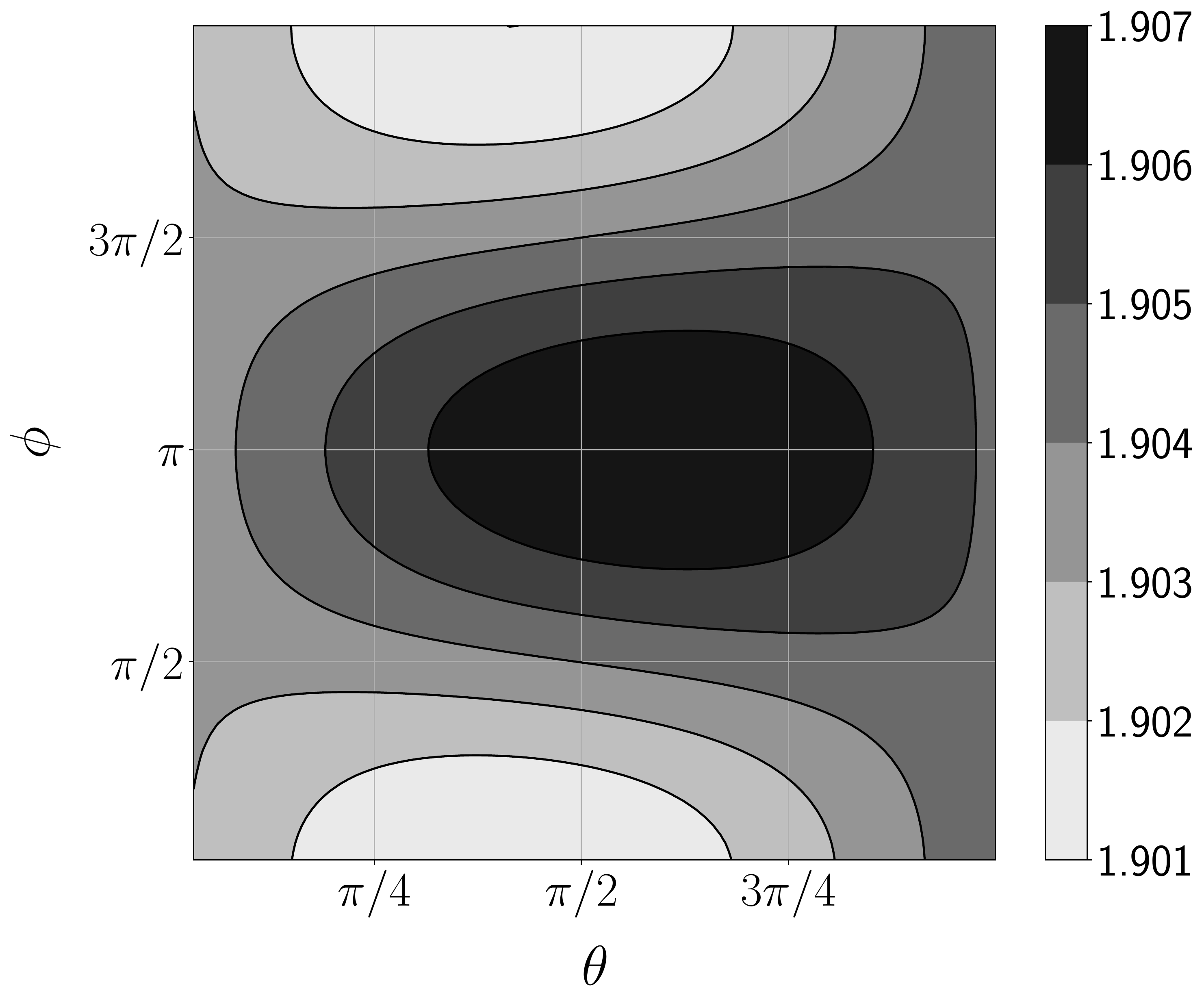} 
}\hfill
\subfloat{
\includegraphics[width=1.0\columnwidth]{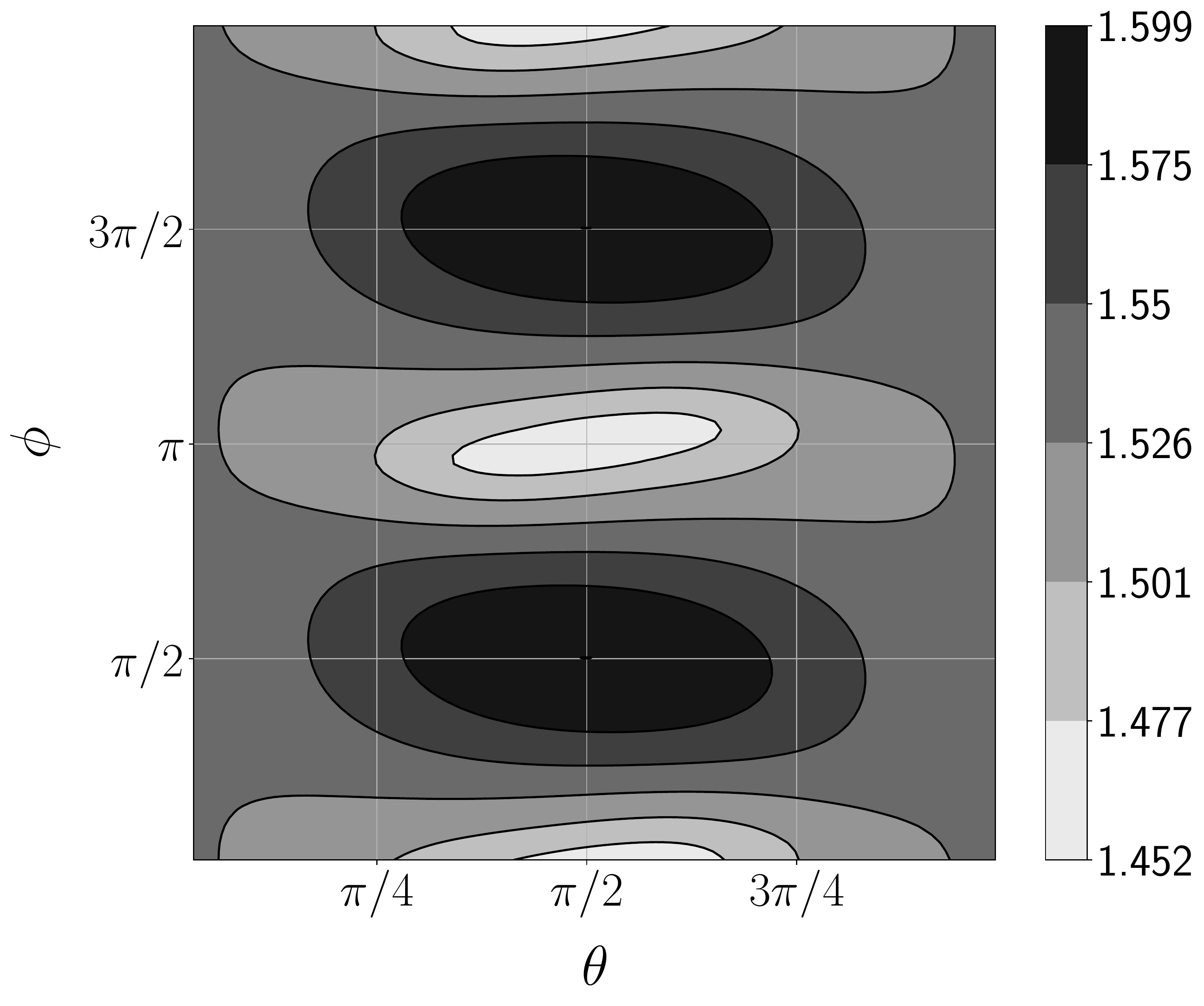} 
}
\caption{Error of the estimation procedure, $\Delta(\mathcal{G},\boldsymbol{s})$, for the quadrant (left) and continuous (right) approaches to the estimation procedure as a function of the parameters defining a pure initial spin state, $\theta$ and $\phi$. 
The setup parameters used were $g_1=2.0$, $g_2=3.24$, $\lambda=0.3$. and $T=1.0$.}
\label{fig:error_initial_state}
\end{figure*}

Since the logarithmic error depends on seven parameters, a relatively large parameter space, we need to focus on a sensible parameter subspace. 
We will consider initial pure spin states, which can be parametrized by  the angles $\theta \in \left[0,\pi\right]$ and $\phi \in \left(0,2\pi\right]$, where $s_1=\sin\theta\cos\phi$, $s_2=\sin\theta\sin\phi$ and $s_3=\cos\theta$.
We will assume no free evolution after the beam interacts with the magnetic field; that is, $T=1$.
In the usual setup of the Stern-Gerlach experiment, the additional free evolution helps to clearly split the beam, guaranteeing a projective measurement of the spin component in that direction. 
Here, no beam separation is expected; therefore, this free evolution is not necessary.
However, the influence of the parameter $T$ will be considered at the end of this section. 
In previous studies~\cite{Potel2005,Hsu2011},  the deflection of the beam in the usual experimental setup was found to be sizable when the product $g_1g_2$ exceeds unity.
We will consider values of $g_1\in [1.0,5.0]$ and $g_2\in (0,4.0]$.
These values for $g_1$ and $g_2$, similar to those used in these studies, are far from the usual approximation where $g_1\gg g_2$~\cite{Potel2005}.

The calculate $\Delta(\mathcal{G},\boldsymbol{s})$ we must numerically determine functions $M_\mu(x,z,T)$ and components $M_{k\mu}(T)$. For this purpose, we use a numerical method based on the Trotter-Suzuki expansion (see Appendix~\ref{sec:app3}). In the following computations, the $x$ and $z$ coordinates are sampled over the interval $[-50,50]$, the total number of samples in each direction is $N_x=N_z=600$, and the total number of temporal steps is $N_t=600$.

Since the exploration of the reduced parameter space would be quite time consuming, we consider the variation of $\Delta(\mathcal{G},\boldsymbol{s})$ as a function of $g_1$ and $g_2$ for different values of $\lambda$ and a fixed initial spin state, as shown in Fig.~\ref{fig:trace_distance_th} for the linear inversion and discrete maximum-likelihood estimators and in Fig.~\ref{fig:trace_distance_th_cont} for the continuous maximum-likelihood estimator.
The error for the chosen initial state, defined by the values $\theta=1.91$ and $\phi=4.78$, is maximum in a setup where $g_1=4.0$, $g_2=0.4$ and $\lambda=0.3$ when using the quadrant approach to the estimation procedure. 
We expect this error to be a pessimistic estimation of the typical error for other values of the parameters  $g_1$, $g_2$ and $\lambda$, and for the continuous maximum-likelihood estimation procedure.
We choose values of $\lambda$ for which the error shows local minima.

Inspection of Figs.~\ref{fig:trace_distance_th} and~\ref{fig:trace_distance_th_cont} shows that the error $\Delta(\mathcal{G},\boldsymbol{s})$ greatly decreases when using the continuous maximum-likelihood estimator instead of the linear inversion or discrete maximum-likelihood estimators. When using a quadrant approach to the estimation procedure, there are regions of the parameter space where the error sharply increases and others where it is relatively low. For the continuous distribution approach, on the other hand, the error remains lower and stabler.

To test our suspicion that the variance of $\check{s}_2$ is responsible for large errors, we plot in Figs.~\ref{fig:trace_distance_tw} and~\ref{fig:trace_distance_tw_cont} the error of the estimation excluding this variance. We find that for most setups this is indeed the case when using a quadrant approach for estimation of the spin state. For example, for a setup defined by $g_1=2.2$, $g_2=2.4$ and $\lambda=1.1$, the variance of $\check{s}_2$ is of order of $10^{12}$, while the variances of $\check{s}_1$ and $\check{s}_3$ are $3.82$ and $2.64$, respectively; for $g_1=4.6$, $g_2=0.24$ and $\lambda=0.2$, the variance of $\check{s}_2$ is of order of $10^8$, while the variances of $\check{s}_1$ and $\check{s}_3$ are $4.15$ and $34.2$, respectively. Notice that this behaviour is found not only for values of $\lambda$ close to unity. However, there are setups where the variance of $\check{s}_2$ is not the largest one. For example, for or $g_1 = 1.0$, $g_2 = 3.96$ and $\lambda = 0.2$, the variances of
$\check{s}_1$, $\check{s}_2$, and $\check{s}_3$ are $10.6$, $60.8$, and $119.0$, respectively.

When using the continuous distribution approach, the error of the estimation behaves similarly, although the variances of $\check{s}_2$ are considerably lower. For $g_1=4.0$, $g_2=3.1$ and $\lambda=1.1$ the variance of $\check{s}_2$ is of order of $10^3$, while the variances of $\check{s}_1$ y $\check{s}_3$ are 2.18 and 1.82, respectively; for $g_1=3.3$, $g_2=0.19$ and $\lambda=1.1$ the variance of $\check{s}_2$ is of order of $10^4$, while the variances of $\check{s}_1$ y $\check{s}_3$ are 7.68 and 7.07, respectively. We also find cases where the variance of $\check{s}_2$ is not the largest one: for $g_1=4.9$, $g_2=0.37$ and $\lambda=0.2$ the variances of $\check{s}_1$, $\check{s}_2$, and $\check{s}_3$ are $1.68$, $3.10$, and $4.26$, respectively.

\begin{figure*}[htbp!]
\centering
\subfloat{
\includegraphics[width=1.0\columnwidth]{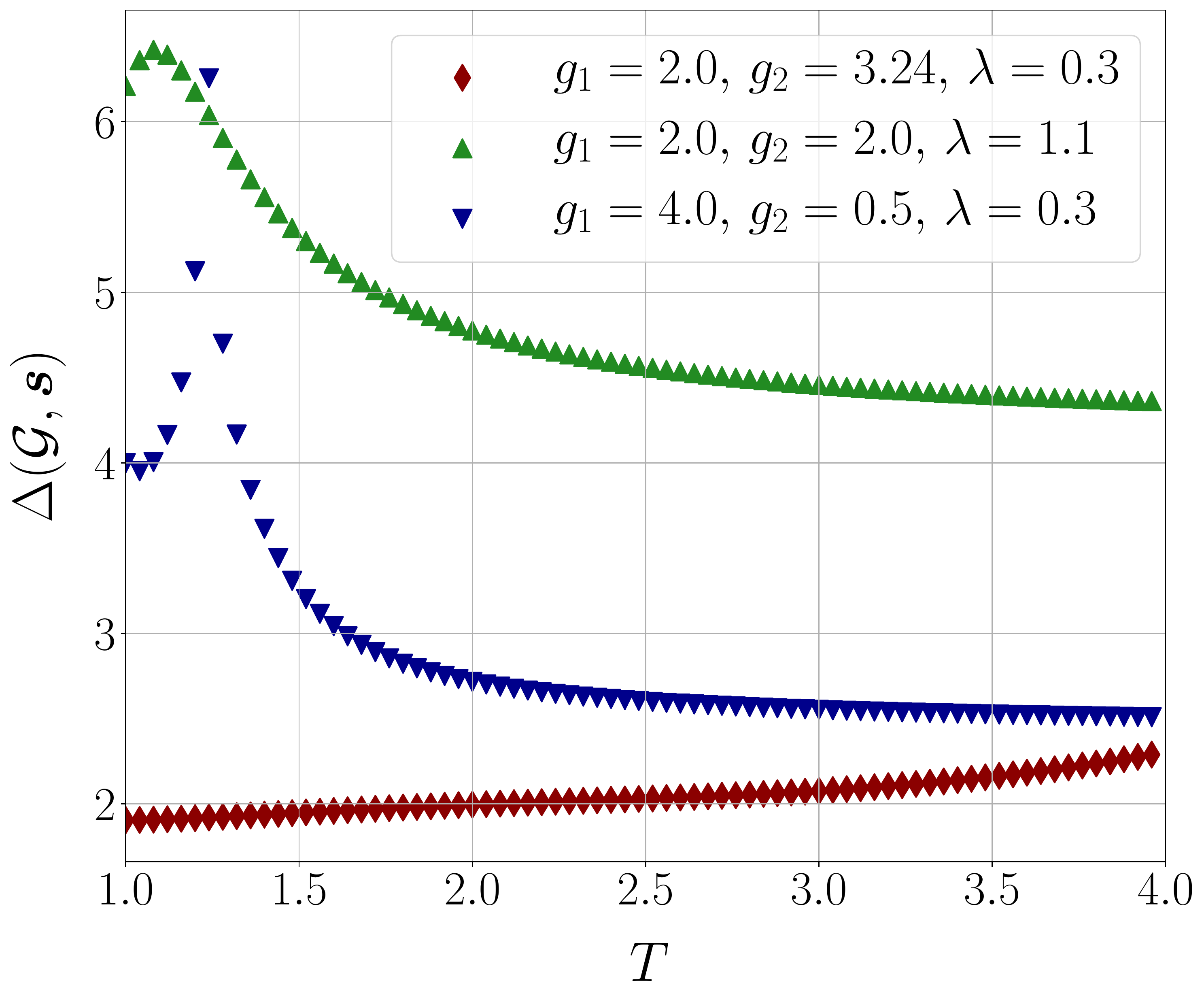} 
}\hfill
\subfloat{
\includegraphics[width=1.0\columnwidth]{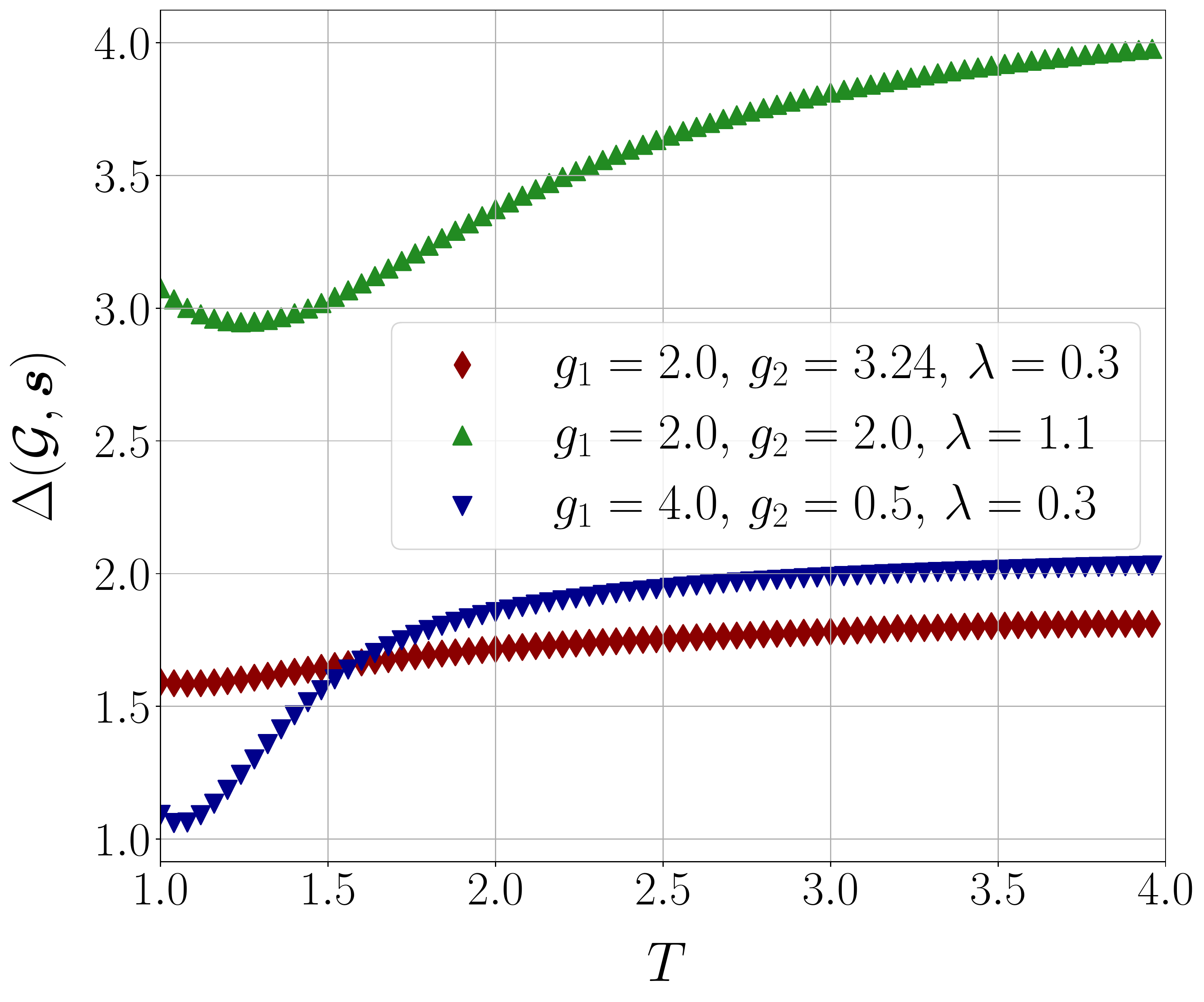} 
}
\caption{Error of the estimation procedure, $\Delta(\mathcal{G},\boldsymbol{s})$, for the quadrant (left) and continuous (right) approaches to the estimation procedure as a function of the detection time, $T$, for the setups $g_1=2.0$, $g_2=3.24$, $\lambda=0.3$ (red, diamond), $g_1=2.0$, $g_2=2.0$, $\lambda=1.1$ (green, triangle up) and $g_1=4.0$, $g_2=0.5$, $\lambda=0.3$ (blue, triangle down). The initial spin state of the beam was a pure state defined by the values $\theta=1.91$ and $\phi=4.78$.}
\label{fig:error_free_evolution}
\end{figure*}

Large variances associated to the estimation of one of the parameters $s_\mu$ indicate that the corresponding probability distribution or probability density function encodes very little information about this parameter. From this point of view, we see that the multinomial distribution~\eqref{eq:multinomial_dist} generally contains less information about parameter $s_2$ than the probability density function~\eqref{eq:cont_density_function}. This is also true for parameters $s_1$ and $s_3$. Thus, it is advisable to use an estimation procedure that uses the complete intensity distribution on the detection screen in order to obtain a better state estimation.

The best regions to perform state estimation are those where the error remains low and stable under small, but not infinitesimal, changes of the parameters that define the experimental setup. For the quadrant approach, for example, when $g_1=2.0$, $g_2=3.24$, $\lambda=0.3$, the variance associated to $\check{s}_2$ is $55.3$, while those associated to $\check{s}_1$ and $\check{s}_3$ are $3.75$ and $2.12$, respectively. For the continuous distribution approach, when the setup is defined by $g_1=4.9$, $g_2=0.37$ and $\lambda=0.2$, the variances for $\check{s}_1$, $\check{s}_2$ and $\check{s}_3$ are   
$1.68$, $3.10$ and $4.26$, respectively.

For a given set of parameters $g_1$, $g_2$ and $\lambda$, the error of the estimation depends on the spin state to be estimated. 
However, if the difference between the lowest and the largest possible error remains sufficiently small, as in the examples of Fig.~\ref{fig:error_initial_state},
the error of the estimation procedure can be defined as the error associated to the state with the worst possible estimation.

To quantify the role of the free evolution on the estimation error, it is necessary to increase the $(x,z)$ region where the intensity distribution is calculated because the wavefunction broadens.
For this computation, the $x$ and $z$ coordinates were sampled over the interval $[-100,100]$, and the total number of samples in each direction was increased to $650$. 

In Fig.~\ref{fig:error_free_evolution} we show examples of the influence of the parameter $T$ on the estimation error. For the quadrant approach, there are cases where the error monotonically grows with $T$ (for example, for $g_1=2.0$, $g_2=3.24$, and $\lambda=0.3$); for other setups, it rapidly increases before decreasing again and reaching a stable value, lower than the one obtained just after the interaction with the magnetic field (for example, for $g_1=4.0$, $g_2=0.5$, and $\lambda=0.3$). 
However, errors are larger than the minimum error found without free evolution. For the continuous distribution approach, we also find setups with monotonic growth of error (for example, for $g_1=2.0$, $g_2=3.24$, and $\lambda=0.3$). However, unlike the previous case, we find setups where the error decreases to a minimum before increasing again and stabilizing (for example, for $g_1=4.0$, $g_2=0.5$, and $\lambda=0.3$). These errors are also usually larger than the minimum error found without free evolution.

Even if it is not generally the case, having a setup where the time of detection ensures a lower value of the error of estimation could prove useful for experimental situations where optimal values for parameters $g_1$ and $g_2$ cannot be easily obtained.
In these situations, one could choose an optimum value for $\lambda$ and/or other regions over which the intensity is evaluated (as a modification to the quadrant approach), to lower the error as much as possible.

As a final remark, we would like to compare the variances obtained by our estimation procedures with the lowest possible values they can take. These lowest possible values are determined by the quantum Cramér-Rao lower bounds~\cite{Watanabe2013,Helstrom1967}, which state the lower possible values the information matrix $\mathbb{J}^{-1}(\boldsymbol{s})$ can take:
\begin{equation}
\mathbb{J}^{-1}(\boldsymbol{s})\geq\frac{1}{N}[\mathbb{J}^{(S)}(\boldsymbol{s})]^{-1},\,\,\mathbb{J}^{-1}(\boldsymbol{s})\geq\frac{1}{N}[\mathbb{J}^{(R)}(\boldsymbol{s})]^{-1}.    
\label{eq:quantum_cramer_rao}
\end{equation}
The matrices $\mathbb{J}^{(S)}(\boldsymbol{s})$ and $\mathbb{J}^{(R)}(\boldsymbol{s})$ are called the symmetric and right \textit{quantum Fisher information matrices}~\cite{Watanabe2013}. These matrices are interpreted as a maximization of the information matrix over all the POVMs chosen for the estimation of the initial spin state. Since the POVM used in this work represents the whole Stern-Gerlach setup, including intensity measurement and spatial state preparation, $\mathbb{J}^{(S)}(\boldsymbol{s})$ and $\mathbb{J}^{(R)}(\boldsymbol{s})$ are interpreted as a maximization of $\mathbb{J}(\boldsymbol{s})$ over the set of parameters $\mathcal{G}=\{g_1,g_2,\lambda,T\}$ and over all the possible forms of measuring the final intensity distribution of the beam. Thus, the quantum Fisher information matrices will only depend on the initial spin state to be estimated.

The components of $\mathbb{J}^{(S)}(\boldsymbol{s})$ and $\mathbb{J}^{(R)}(\boldsymbol{s})$ are calculated as
\begin{equation}
[\mathbb{J}^{(S)}(\boldsymbol{s})]_{\mu\nu}=\frac{1}{2}\left\langle\left\{L_\mu,L_\nu\right\}\right\rangle,
\label{eq:symmetric_fisher_comp}
\end{equation}
\begin{equation}
[\mathbb{J}^{(R)}(\boldsymbol{s})]_{\mu\nu}=\left\langle L'_\nu L'_\mu\right\rangle,
\label{eq:right_fisher_comp}
\end{equation}
where the symbol $\{\cdot\}$ indicates an anticommutator and $\langle\cdot\rangle=\textrm{Tr}_S[(\cdot)\rho_S(0)]$. Operators $L_\mu$ and $L'_\mu$ are called the symmetric and right \textit{logarithmic derivatives}, respectively and are defined as~\cite{Watanabe2013}
\begin{equation}
\frac{\partial\rho_S(0)}{\partial s_\mu}=\frac{1}{2}\left\{\rho_S(0),L_\mu\right\}\,\,,\,\,\frac{\partial\rho_S(0)}{\partial s_\mu}=\rho_S(0)L'_\mu.
\label{eq:logarithmic_derivatives}
\end{equation}
From these relations, it can be shown that
\begin{equation}
[\mathbb{J}^{(S)}(\boldsymbol{s})]_{\mu\nu}^{-1}=\frac{1}{2}\langle\{\sigma_\mu,\sigma_\nu\}\rangle-\langle\sigma_\nu\rangle\langle\sigma_\nu\rangle,
\label{eq:symmetric_inv_fisher_comp}
\end{equation}
\begin{equation}
[\mathbb{J}^{(R)}(\boldsymbol{s})]_{\mu\nu}^{-1}=\langle\sigma_\mu\sigma_\nu\rangle-\langle\sigma_\mu\rangle\langle\sigma_\nu\rangle,
\label{eq:right_inv_fisher_comp}
\end{equation}
which correspond to the symmetrized and unsymmetrized covariance matrices of the Pauli spin operators, respectively~\cite{Watanabe2011}. 

The previous treatment is only valid for a mixed initial spin state. For a pure initial spin states, it can be shown that $[\mathbb{J}^{(S)}(\boldsymbol{s})]_{\mu\nu}^{-1}=\delta_{\mu\nu}$~\cite{Fujiwara1995}. However, operators $L_\mu'$ cannot be defined and the components $[\mathbb{J}^{(R)}(\boldsymbol{s})]_{\mu\nu}^{-1}$ are not easily calculated~\cite{FujiwaraRLD,Fujiwara1999}. As a consequence, we will take into account only $[\mathbb{J}^{(S)}(\boldsymbol{s})]^{-1}$ as the reference for the lower bound of the information matrix for the case of pure initial spin states.

In analogy to equation~\eqref{eq:quality_measure}, we define the logarithmic quantum error of the estimation as 
\begin{equation}
\Delta_Q(\boldsymbol{s})=\log_{10}\left[\textrm{tr}\left([\mathbb{J}^{(Q)}(\boldsymbol{s})]^{-1}\right)\right]\,\,,\,\,Q=S,R,
\label{eq:quantum_quality_measure}
\end{equation}
which satisfies the relation $\Delta(\mathcal{G},\boldsymbol{s})\geq\Delta_Q(\boldsymbol{s})$. Using Eqs.~\eqref{eq:symmetric_inv_fisher_comp} and~\eqref{eq:right_inv_fisher_comp}, we see that \begin{equation*}
\Delta_S(\boldsymbol{s})=\Delta_R(\boldsymbol{s})=\log_{10}\left(3-\sum_{\mu=1}^3s_\mu^2\right)
\end{equation*}
for mixed initial spin states. Since $\sum_{\mu=1}^3s_\mu^2<1$, $\Delta(\mathcal{G},\boldsymbol{s})>\log_{10}(2)\approx0.301$. For pure spin states $\Delta(\mathcal{G},\boldsymbol{s})\geq\Delta_S(\boldsymbol{s})=\log_{10}(3)\approx0.477$.

As can be seen in Figs.~\ref{fig:trace_distance_th} to~\ref{fig:error_free_evolution}, our estimation procedure does not attain the lowest possible bound. 
In the explored region of parameters, only the estimations of parameters $s_1$ and $s_3$ are close to the optimal value of the error, when using a continuous distribution approach to the state estimation.
However, in the case of a real experiment, the suitable choice of the number of particles can help to obtain reasonable values for the variances of all the parameters that define the initial spin state. 

\section{\label{sec:conc}Conclusions and perspectives}
In this work we have shown how a modified setup of the Stern-Gerlach experiment can be used to estimate the initial spin state of a beam of neutral spin-1/2 particles. 
There are three modifications: the use of a magnetic field without a large reference component, the measurement of the spatial intensity distribution of the beam over at least four different regions or over the complete plane of detection, and the suitable choice of the initial spatial state of the beam of particles.

Using a quantum-mechanical description of the experimental setup, we derived linear inversion and  maximum-likelihood estimation procedures for the parameters that define the initial spin state. It was found that,
unless the initial spin state is rotationally invariant along the direction of propagation of the beam, all of the parameters that define the initial spin state can be estimated.

The quality of the estimation of the initial spin state was quantified by the logarithm of the sum of the variances of the parameters which characterize the state (Bloch vector components).
This measure allowed us to compare the errors associated to different experimental setups and to find the typical values of the variances that can be obtained with the use of the estimation procedures. 
Although these variances do not generally attain the lower limit imposed by quantum Cramér-Rao bound, they can take reasonably low values when the number of particles of the beam is large enough. An optimization of the error of estimation could reveal possible experimental setups that attain variances that are closer to this lower bound.

A straightforward rotation of the usual experimental setup allows the measurement of the spin components transverse to the propagation direction of the beam.
However, we are not aware of any setup for neutral beams, which enables the estimation of the spin component in the direction of propagation without previously changing the spin of the particles (by means of another magnetic field, for example).
Thus, despite its shortcomings and technical difficulties, the proposal made in this paper may be a viable alternative to estimate the spin state of neutral beams.

It is interesting to discuss a possible set of experimental parameters compatible with the values of $g_1$ and $g_2$ that we chose for the quantification of the estimation error. 
In terms of the real experimental parameters, $g_1=\mu b\sigma\tau/2\hbar$, and $g_2=\hbar\tau/2m\sigma^2$.  
We will assume that the particles of the beam are neutrons, in this way we fix the values of $\mu$ and $m$ to $\mu=0.97\times10^{-26}\,\textrm{J/T}$ and $m=1.67\times10^{-27}\,\textrm{kg}$. 
Usual field gradients in Stern-Gerlach experiments 
vary between $1\,\textrm{T/m}$ and $100\,\textrm{T/m}$~\cite{Jones1980,Sherwood1954,Hamelin1975}. 
If the neutrons are slow enough, a large gradient is not necessary, so it is reasonable to assume that $b\sim1\,\textrm{T/m}$. 
In these same experiments, the length of the magnet is usually close to $1\,\textrm{m}$; we will take this value as a reasonable length for the magnet. 
Experiments with cold neutrons report average beam speeds between $400\,\textrm{m/s}$ and $600\,\textrm{m/s}$~\cite{Jones1980}. 
Assuming these speeds, the time of interaction with the magnetic field would vary between $\tau\sim1.7\,\textrm{ms}$ and $\tau\sim2.5\,\textrm{ms}$.
By taking these values for $b$ and $\tau$, and considering the conditions over $g_1$ and $g_2$ that were used to calculate the error of the estimation, $\sigma$ would vary between $\sigma\sim5\,\mu\textrm{m}$ and $\sigma\sim10\,\mu\textrm{m}$. 

We consider the values for speeds, field gradients, and other physical quantities discussed on the previous paragraph, to be adequate for an experimental implementation of the estimation procedure.
Although actual experimental results might significantly differ from our numerical results, due to the idealizations we have made in the model Hamiltonian (like neglecting the variation of the magnetic field along the direction of the beam), we would expect state estimation to be possible.


\appendix

\section{\label{sec:app1}Intensity measurements for $\lambda=1$}
In this appendix we show that the spatial intensity distribution of the beam does not encode information about the parameter $s_2$ when $\lambda=1$.

The Hamiltonian $H_{xz}$ is, in polar coordinates
\begin{equation}
 H_{xz}=g_2\left(p_r^2+\frac{L_y^2}{r^2}\right)+g_1r\left(\cos\theta\sigma_1-\sin\theta\sigma_3\right),
\label{eq:polar_coord_hamiltonian}
\end{equation}
where $p_r$ is the radial momentum, $L_y$ the angular momentum in the $y$ direction, $x=r\cos\theta$, and $z=r\sin\theta$.
The initial spatial state, expressed in the same coordinates, is
\begin{align}
\begin{split}
\braket{r,\theta|\psi_{xz}} = 
\sqrt{\frac{1}{2\pi\lambda}}
e^{-\frac{r^2}{4}} 
\exp\left[\left(\lambda^2-1\right)\frac{r^2\sin^2\theta}{4\lambda^2}\right].    
\end{split}
\end{align} 

While the state of the beam at time of detection is $\rho(T)=U_{xz}(T,0)|\psi_{xz}\rangle\langle\psi_{xz}|\rho_S(0)U_{xz}^\dagger(T,0)$, the evolution operator can be factorized as $U_{xz}(T,0)=U_{xz}^{(f)}(T,1)U_{xz} (1,0)$.
By expanding $U_{xz} (1,0)$ as $U_{xz} (1,0)=\sum_{\alpha=0}^3A_\alpha\sigma_\alpha$, and the initial spin state as $\rho_S(0)=(1/2)\sum_{\mu=0}^3s_\mu\sigma_\mu$, we find
\begin{equation}
\rho(T)=\frac{1}{2}\sum_{\alpha,\beta,\mu=0}^3\sigma_\alpha\sigma_\mu\sigma_\beta|\phi_\alpha(T)\rangle\langle\phi_\beta(T)|s_\mu,
\label{eq:state_Pauli_app}
\end{equation}
where $|\phi_\alpha(T)\rangle=U_{xz}^{(f)}(T,1)A_\alpha|\psi_{xz}\rangle$.

We expand operator $U_{xz} (1,0)$ in a power series of the Hamiltonian $H_{xz}$, 
$U_{xz} (1,0) = \sum_{k=0}^{\infty}\frac{(-i)^k}{k!}H_{xz}^k$.
We also expand each power of the Hamiltonian as $H_{xz}^k=\sum_{\alpha=0}^3h_\alpha^{(k)}\sigma_\alpha$ where $\{h_\alpha^{(k)}\}$ are spatial Hermitian operators. 
In this way, 
$U_{xz}(1,0)=\sum_{\alpha=0}^3\sum_{k=0}^{\infty}\frac{(-i)^k}{k!}h_\alpha^{(k)}\sigma_\alpha.$
By direct comparison, $A_\alpha$ is found to be
\begin{equation}
A_\alpha=\sum_{k=0}^{\infty}\frac{(-i)^k}{k!}h_\alpha^{(k)}.
\label{eq:time_evolution_pauli_recursion}
\end{equation}
Since the coefficients $\{h_\alpha^{(k)}\}$ are obtained from powers of the Hamiltonian, we can find recurrence relations between them for each order in the power series. 
By using relation $H_{xz}^{(k+1)}=H_{xz}H_{xz}^{(k)}$, we find the following expressions for the computation of the coefficients at higher orders:
\begin{equation}
h_0^{(k+1)}=g_2P^2h_0^{(k)}+g_1r\left(\cos\theta h_1^{(k)}-\sin\theta h_3^{(k)}\right),
\label{eq:recurrence_relation_0}
\end{equation}
\begin{equation}
h_1^{(k+1)}=g_2P^2h_1^{(k)}+g_1r\left(\cos\theta h_0^{(k)}+i\sin\theta h_2^{(k)}\right),
\label{eq:recurrence_relation_1}
\end{equation}
\begin{equation}
h_2^{(k+1)}=g_2P^2h_2^{(k)}-ig_1r\left(\cos\theta h_3^{(k)}+\sin\theta h_1^{(k)}\right),
\label{eq:recurrence_relation_2}
\end{equation}
\begin{equation}
h_3^{(k+1)}=g_2P^2h_3^{(k)}-g_1r\left(\sin\theta h_0^{(k)}-i\cos\theta h_2^{(k)}\right),
\label{eq:recurrence_relation_3}
\end{equation}
where we have made the definition $P^2=p_r^2+r^{-2}L_y^2$. 
These relations are complemented by the initial conditions $h_0^{(0)}=I_{xz}$, $h_1^{(0)}=h_2^{(0)}=h_3^{(0)}=0$, where $I_{xz}$ is the identity operator over $\mathcal{H}_{xz}$.

When acting over the initial spatial state, coefficients $\{h_\alpha^{(k)}\}$ satisfy the following relations for every order: 
\begin{align}
&h_2^{(k)}|\psi_{xz}\rangle = (1-\lambda^2)\,G_2^{(k)}|\psi_{xz}\rangle,
\label{eq:recurr_property1}    
\\ &
\left(\sin\theta\,h_1^{(k)}+\cos\theta\,h_3^{(k)}\right)|\psi_{xz}\rangle= 
(1-\lambda^2)\,G_{13}^{(k)}|\psi_{xz}\rangle, 
\label{eq:recurr_property2}
\\ &
\left(\sin\theta L_yh_3^{(k)}-\cos\theta L_yh_1^{(k)}\right)|\psi_{xz}\rangle= 
(1-\lambda^2)\,F_{13}^{(k)}|\psi_{xz}\rangle, 
\label{eq:recurr_property3}    
\\ &
L_yh_0^{(k)}|\psi_{xz}\rangle = (1-\lambda^2)\,G_0^{(k)}|\psi_{xz}\rangle. 
\label{eq:recurr_property4}    
\end{align}
Operators $G_2^{(k)}$, $G_{13}^{(k)}$, $F_{13}^{(k)}$ and $G_{0}^{(k)}$ generally depend on $r$, $\theta$ and $\lambda$. 

To prove these properties, we will proceed by induction. 
At first order, these properties are valid; there are two non-vanishing terms, $F_{13}^{(1)}=\frac{i\,g_1r^3\sin\left(2\theta\right)}{4\lambda^2}$ and 
\begin{align*}
\begin{split}
G_{0}^{(1)}=\frac{-ig_2r^2\sin(2\theta)}{32\lambda^6}&\left[(\lambda^4-1)r^2\cos(2\theta)\right.\\
&\left.+(\lambda^4+1)r^2-12\lambda^2(\lambda^2+1)\right].
\end{split}    
\end{align*}
Assuming that all properties hold at order $k$, we obtain the following expressions for the operators at order $k+1$:
\begin{align*}
G_{2}^{(k+1)}&=g_2P^2G_2^{(k)}-ig_1rG_{13}^{(k)},    
\\
G_{13}^{(k+1)}&=g_2P^2G_{13}^{(k)}-g_2r^{-2}\left(G_{13}^{(k)}+2iF_{13}^{(k)}\right)+ig_1rG_2^{(k)},   
\\
F_{13}^{(k+1)}&=g_2\left(P^2+3r^{-2}\right)F_{13}^{(k)}-g_1r\left(G_0^{(k)}+G_2^{(k)}\right)\\
&\qquad+2ig_2r^{-2}\left(L_y^2-1\right)G_{13}^{(k)},    
\\
G_{0}^{(k+1)} & =g_2P^2G_0^{(k)}-g_1r\left(F_{13}^{(k)}-i\,G_{13}^{(k)}\right).
\end{align*}
Therefore, relations~\eqref{eq:recurr_property1} to~\eqref{eq:recurr_property4} hold for every order. 


We use now Eq.~\eqref{eq:time_evolution_pauli_recursion} to express the previous properties in terms of operators $\{A_\alpha\}$:
\begin{equation}
A_2|\psi_{xz}\rangle = (1-\lambda^2)\,G_2|\psi_{xz}\rangle,
\label{eq:property1}    
\end{equation}
\begin{align}
\begin{split}
\left(zA_1+xA_3\right)|\psi_{xz}\rangle=(1-\lambda^2)\,G_{13}|\psi_{xz}\rangle,
\end{split}
\label{eq:property2}
\end{align}
\begin{align}
\begin{split}
\left(xL_yA_1-zL_yA_3\right)|\psi_{xz}\rangle=(1-\lambda^2)\,F_{13}|\psi_{xz}\rangle,
\end{split}
\label{eq:property3}    
\end{align}
\begin{equation}
L_yA_0|\psi_{xz}\rangle = (1-\lambda^2)\,G_0|\psi_{xz}\rangle,
\label{eq:property4}    
\end{equation}
where $G_2$, $G_{13}$, $F_{13}$ and $G_{0}$ are obtained from the corresponding series of operators $\{G_2^{(k)}\}$, $\{G_{13}^{(k)}\}$, $\{F_{13}^{(k)}\}$ and $\{G_{0}^{(k)}\}$, respectively.  

Now we can explore the implications of having $\lambda=1$. Eq.~\eqref{eq:property1} implies that $|\phi_2(T)\rangle=0$.
Eq.~\eqref{eq:property2}, on the other hand, implies that $A_1|\psi_{xz}\rangle=xA|\psi_{xz}\rangle$ and $A_3|\psi_{xz}\rangle=-zA|\psi_{xz}\rangle$. Additionally, when combined with Eq.~\eqref{eq:property3}, yields to the relation $(p_zA_1+p_xA_3)|\psi_{xz}\rangle=0$, which allows to see that
\begin{align}
\begin{split}
\left(z\,U_{xz}^{(f)}(T,1)A_1+x\,U_{xz}^{(f)}(T,1)A_3\right)|\psi_{xz}\rangle=0.
\end{split}
\label{eq:property2T}
\end{align}
This means that $x|\phi_3(T)\rangle=-z|\phi_1(T)\rangle$, which, in turn, implies that $|\phi_3(T)\rangle\langle\phi_1(T)|-|\phi_1(T)\rangle\langle\phi_3(T)|=0$.

These results have an enormous influence in the structure of the spatial intensity distribution of the beam. 
Remembering the expression $I(x,z) = \operatorname{Tr}_S\left(\braket{x,z|\rho(T)|x,z}\right)$ and using Eq.~\eqref{eq:state_Pauli_app}, we see that  
\begin{equation}
I(x,z)=\sum_{\alpha,\beta,\mu=0}^3d_{\alpha\mu\beta}\phi_\alpha(x,z,T)\phi_\beta^{*}(x,z,T)s_\mu,
\label{eq:intensity_Pauli_app}
\end{equation}
where $d_{\alpha\mu\beta}=\textrm{Tr}_S(\sigma_\alpha\sigma_\mu\sigma_\beta)/2$ and functions $\phi_\alpha(x,z,T)$ are calculated as $\langle x,z|\phi_{\alpha}(T)\rangle$. 
For the intensity distribution to depend on $s_2$, the term $2\textrm{Re}\left(\phi_0\phi_2^{*}\right)-2\textrm{Im}\left(\phi_1\phi_3^{*}\right)$ must be different from zero. 
However, when $\lambda=1$, this term identically vanishes, and thus, the estimation of $s_2$ cannot be achieved by using intensity measurements over any region of the $(x,z)$ plane.

\section{\label{sec:app2}Computation of the maximum-likelihood estimators}
In this appendix we will show how to derive the maximum-likelihood estimators for the initial spin state.

In the case of the quadrant approach, the log-likelihood function reads
\begin{equation}
l(\boldsymbol{s}|\boldsymbol{n})=\ln N!-\sum_{k=1}^4n_k! + N\sum_{k=1}^4f_k\ln [p_k(T)].
\label{eq:log_likelihood}
\end{equation}
Taking the variation of this function with respect to $\rho_S(0)$ and remembering that $p_k(T)=\textrm{Tr}_S[\tilde{Q}_k(T)\rho_S(0)]$ we see that
\begin{align}
\begin{split}
\delta l(\boldsymbol{s}|\boldsymbol{n})&=N\sum_{k=1}^4f_k\delta\ln [p_k(T)]\\ 
&=N\sum_{k=1}^4\frac{f_k}{p_k}\textrm{Tr}_S[\tilde{Q}_k(T)\delta\rho_S(0)]\\
&=N\textrm{Tr}_S\left[\left(\sum_{k=1}^4\frac{f_k}{p_k}\tilde{Q}_k(T)\right)\delta\rho_S(0)\right]\\
&=N\textrm{Tr}_S[R_S\delta\rho_S(0)],
\end{split}
\label{eq:variation_log_likelihood}
\end{align}
where $$R_S=\sum_{k=1}^4\frac{f_k}{p_k}\tilde{Q}_k(T).$$ In this way, a maximum-likelihood estimator for the spin state, which we will denote $\check{\rho}_S$, must satisfy the relation
\begin{equation}
\left.\textrm{Tr}_S[R_S\delta\rho_S(0)]\right|_{\rho_S(0)=\check{\rho}_S}=0.
\label{eq:maximal_equation}    
\end{equation}

Now we must determine the variation of the $\rho_S(0)$. Here is where the different constraints over the estimation of the spin state can be included. If we ask for the estimated state to be only normalized, we can express the initial spin state as
\begin{equation*}
\rho_S(0)=\frac{\beta}{\textrm{Tr}_S(\beta)},    
\end{equation*}
where $\beta$ is an arbitrary Hermitian operator. This form of $\rho_S(0)$ implies that
$$\delta\rho_S(0)=\frac{\delta\beta}{\textrm{Tr}_S(\beta)}-\rho_S(0)\frac{\textrm{Tr}_S(\delta\beta)}{\textrm{Tr}_S(\beta)}.$$
Replacing into Eq.~\eqref{eq:maximal_equation}, and after some algebra, we find that the estimator of the spin state must satisfy the relation
\begin{equation}
\left.\frac{1}{\textrm{Tr}_S(\beta)}\textrm{Tr}_S[(R_S-\sigma_0)\delta\beta]\right|_{\rho_S(0)=\check{\rho}_S}=0.
\label{eq:lin_inv_equation_1}    
\end{equation}
Since $\delta\beta$ is arbitrary, we find
\begin{equation}
\check{R}_S=\sigma_0,
\label{eq:lin_inv_equation_2}    
\end{equation}
where $\check{R}_S$ is operator $R_S$ calculated using values $\check{p}_k(T)=\textrm{Tr}_S[\tilde{Q}_k(T)\check{\rho}_S]$ instead of probabilities $p_k(T)$. This equation is solved for $f_k=\check{p}_k(T)$, that is
\begin{equation}
f_k=\sum_{\mu=0}^3M_{k\mu}\check{s}_\mu,
\label{eq:lin_inv_equation_3}    
\end{equation}
where we have used the expansion $\check{\rho}_S=\frac{1}{2}\sum_{\mu=0}^3\check{s}_\mu\sigma_\mu$. This is the expression for the linear inversion estimator.

Additionally asking for the state to be positive semidefinite, we can express $\rho_S(0)$ as~\cite{Siah2015}
$$\rho_S(0)=\frac{A^{\dagger}A}{\textrm{Tr}_S(A^{\dagger}A)},$$
where $A$ is an arbitrary and generally non-Hermitian operator. With this definition, $\delta\rho_S(0)$ reads
$$\delta\rho_S(0)=\frac{\delta A^{\dagger}A+A^{\dagger}\delta A}{\textrm{Tr}_S(A^{\dagger}A)}-\frac{\textrm{Tr}_S(\delta A^{\dagger}A+A^{\dagger}\delta A)}{\textrm{Tr}_S(A^{\dagger}A)}\rho_S(0).$$

Replacing into equation~\eqref{eq:maximal_equation}, and after some long algebra, we obtain  
\begin{align}
\begin{split}
&\left.\frac{1}{\textrm{Tr}_S(A^{\dagger}A)}\textrm{Tr}_S[(R_S-\sigma_0)\delta A^{\dagger}A]\right|_{\rho_S(0)=\check{\rho}_S}\\
&-\left.\frac{1}{\textrm{Tr}_S(A^{\dagger}A)}\textrm{Tr}_S[(R_S-\sigma_0) A^{\dagger}\delta A]\right|_{\rho_S(0)=\check{\rho}_S}=0.
\end{split}
\label{eq:ML_disc_equation_1}    
\end{align}
Since $\delta A$ y $\delta A^{\dagger}$ are arbitrary, this relation implies that
\begin{align}
\begin{split}
&\left.A(R_S-\sigma_0)\right|_{\rho_S(0)=\check{\rho}_S}\\
&\;\;\;\;\;\;\;\;\;\;\;\;\;=\left.(R_S-\sigma_0) A^{\dagger}\right|_{\rho_S(0)=\check{\rho}_S}=0
\end{split}
\label{eq:ML_disc_equation_2}    
\end{align}
or, equivalently,
\begin{align}
\begin{split}
&\left.\rho_S(0)(R_S-\sigma_0)\right|_{\rho_S(0)=\check{\rho}_S}\\
&\;\;\;\;\;\;\;\;\;\;\;\;\;=\left.(R_S-\sigma_0) \rho_S(0)\right|_{\rho_S(0)=\check{\rho}_S}=0.
\end{split}
\label{eq:ML_disc_equation_3}    
\end{align}

Thus, the estimator for the initial spin state satisfies the relations
\begin{align}
\begin{split}
\check{R}_S\check{\rho}_S=\check{\rho}_S\check{R}_S=\check{\rho}_S.
\end{split}
\label{eq:ML_disc_equation_4}    
\end{align}

Using $\check{\rho}_S=\frac{1}{2}\sum_{\mu=0}^3\check{s}_\mu\sigma_\mu$, expanding operator $\check{R}_S$ in terms of Pauli spin operators as
$$\check{R}_S=\sum_{\mu=0}^3\check{r}_\mu\sigma_\mu,$$
where 
$$\check{r}_\mu=\sum_{k=1}^4\frac{f_k}{\check{p}_k}M_{k\mu};$$
multiplying Eq.~\eqref{eq:ML_disc_equation_4} by $\sigma_\alpha$ and taking the trace, we find $\check{s}_0=1$ and
\begin{align}
\begin{split}
(1-\check{r}_0)\check{s}_\alpha=\check{r}_\alpha,
\end{split}
\label{eq:ML_disc_equation_5}    
\end{align}
which are the expressions for the discrete maximum-likelihood estimator of the initial spin state.

The procedure to obtain the continuous maximum-likelihood estimator is completely analogous to the one just described. We need only take into account that the log-likelihood is written as 
\begin{equation}
l(\boldsymbol{s}|\boldsymbol{v})=\sum_{k=1}^N\ln [I(x_k,z_k)],
\label{eq:log_likelihood_cont}
\end{equation}
which leads to a change in the definition of operator $R_S$:
$$R_S=\frac{1}{N}\sum_{k=1}^N\frac{\tilde{Q}_{x_kz_k}(T)}{I(x_k,z_k)}.$$
The expansion of operator $\check{R}_S$ in terms of Pauli matrices will then have the form
$$\check{R}_S=\sum_{\mu=0}^3\check{R}_\mu\sigma_\mu,$$
where
$$\check{R}_\mu=\frac{1}{N}\sum_{k=1}^N\frac{M_\mu(x_k,z_k,T)}{\check{I}(x_k,z_k)}$$
with $\check{I}(x,z)=\sum_{\mu=0}^3M_\mu(x,z,T)\check{s}_\mu.$

Thus, the estimators for the initial spin state will read $\check{s}_0=1$ and 
\begin{align}
\begin{split}
(1-\check{R}_0)\check{s}_\alpha=\check{R}_\alpha,
\end{split}
\label{eq:ML_cont_equation}    
\end{align}

Now, for the numerical computation of the maximum-likelihood estimators, we can see that Eq.~\eqref{eq:ML_disc_equation_5} can be cast into the form
\begin{equation}
\check{R}_S\check{\rho}_S\check{R}_S=\check{\rho}_S.
\label{eq:ML_num_1}
\end{equation}
This is the central expression of the $R\rho R$ algorithm~\cite{Rehaceck2007}.

The algorithm states that the estimator $\check{\rho}_S$ can be computed interatively from the relation
\begin{equation}
\check{\rho}_S^{(n+1)}=\mathcal{N}\left[\check{R}_S^{(n)}\check{\rho}_S^{(n)}\check{R}_S^{(n)}\right],
\label{eq:ML_num_2}
\end{equation}
where the symbol $\mathcal{N}[\cdot]$ indicates the normalization to trace one of the corresponding operator. 

Using the expansions of operators $\check{R}_S$ and $\check{\rho}_S$ in terms of Pauli spin matrices, we find for the discrete maximum-likelihood estimator
\begin{equation}
\mathcal{N}\left[\check{R}_S^{(n)}\check{\rho}_S^{(n)}\check{R}_S^{(n)}\right]=2\check{r}_0^{(n)}+\check{\gamma}^{(n)},
\label{eq:ML_num_3}
\end{equation}
where 
$$\check{\gamma}^{(n)}=\sum_{\mu=1}^3\left(\check{r}_\mu^{(n)}\right)^2-\left(\check{r}_0^{(n)}\right)^2.$$
For the continuous maximum-likelihood estimator, we find similar expressions
\begin{equation}
\mathcal{N}\left[\check{R}_S^{(n)}\check{\rho}_S^{(n)}\check{R}_S^{(n)}\right]=2\check{R}_0^{(n)}+\check{\Gamma}^{(n)},
\label{eq:ML_num_4}
\end{equation}
where 
$$\check{\Gamma}^{(n)}=\sum_{\mu=1}^3\left(\check{R}_\mu^{(n)}\right)^2-\left(\check{R}_0^{(n)}\right)^2.$$

Multiplying equation~\eqref{eq:ML_num_2} without the normalization by $\sigma_\alpha$, taking the trace and then dividing by $\mathcal{N}[\check{R}_S^{(n)}\check{\rho}_S^{(n)}\check{R}_S^{(n)}]$, we find following expression for the discrete maximum-likelihood estimator:
\begin{equation}
\check{s}_\alpha^{(n+1)}=\frac{2\check{r}_\alpha^{(n)}-\check{s}_\alpha^{(n)}\check{\gamma}^{(n)}}{2\check{r}_0^{(n)}+\check{\gamma}^{(n)}}.
\label{eq:ML_num_5}
\end{equation}
For the continuous maximum-likelihood estimator, we find a completely analogous expression:
\begin{equation}
\check{s}_\alpha^{(n+1)}=\frac{2\check{R}_\alpha^{(n)}-\check{s}_\alpha^{(n)}\check{\Gamma}^{(n)}}{2\check{R}_0^{(n)}+\check{\Gamma}^{(n)}}.
\label{eq:ML_num_6}
\end{equation}

\section{\label{sec:app3}Numerical calculation of the functions $M_\mu(x,z,T)$}

The complete implementation of the algorithms described by Eqs.~\eqref{eq:iterative_MLE} and~\eqref{eq:iterative_MLE_cont}, and the calculation of the error of the estimation, Eq.~\eqref{eq:quality_measure}, requires the determination of functions $M_\mu(x,z,T)$, which in turn determine the elements of the measurement matrix $M_{k\mu}(T)$. 

As stated before, these functions are the coefficients of the expansion of operator $\tilde{Q}_{xz}(T)=\textrm{Tr}_{xz}\left[U^{\dagger}_{xz}(T,0)Q_{xz}U_{xz}(T,0)R_{xz}(0)\right]$ in terms of Pauli matrices. Recalling that $R_{xz}(0)=|\psi_{xz}\rangle\langle\psi_{xz}|$ and expanding the time evolution operator as $U_{xz}(T,0)=\sum_{\alpha=0}^3A_{\alpha}(T)\sigma_\alpha$, operator $\tilde{Q}_{xz}(T)$ can be recast as 
\begin{equation*}
\tilde{Q}_{xz}(T)=\sum_{\alpha,\beta=0}^3\phi_\alpha(x,z,T)\phi_{\beta}^*(x,z,T)\sigma_\alpha\sigma_\beta,
\end{equation*}
where $\phi_{\alpha}(x,z,T) = \braket{x,z|A_\alpha(T)|\psi_{xz}}$. From this expression we see that
\begin{equation}
M_\mu(x,z,T)=\sum_{\alpha,\beta=0}^3\phi_\alpha(x,z,T)\phi_{\beta}^*(x,z,T)\,d_{\alpha\beta\mu},
\label{eq:func_comp_exp}
\end{equation}
where $d_{\alpha\beta\mu}=\textrm{Tr}_S(\sigma_\alpha\sigma_\beta\sigma_\mu)/2$. Following equation~\eqref{eq:mesaurement_matrix_components}, the components of the measurement matrix are calculated as
\begin{equation}   
M_{k\mu}(T)=\sum_{\alpha,\beta=0}^3d_{\alpha\beta\mu}\int_{\Omega_k}\phi_\alpha(x,z,T)\phi_{\beta}^*(x,z,T)\,dxdz.
\label{eq:matrix_comp_exp}
\end{equation}

As can be seen, the determination of
$M_\mu(x,z,T)$ relies in the computation of the four functions $\phi _{\alpha}(x,z,T)$. These functions are not easily calculated by analytical means, so we will compute them using the numerical method described below. Though we will assume a pure initial spin state of the beam for the description of the method, the results are also valid for mixed initial spin states.

The state of the particles at time $T,$ the time of detection, is $|\psi(T)\rangle= U_{\bar{x}\bar{z}}^{(f)}(T,t) U_{\bar{x}\bar{z}} (t,0) |\psi_{xz}\rangle |\chi\rangle$, where $U_{\bar{x}\bar{z}}^{(f)}(T,t)$ and $U_{\bar{x}\bar{z}} (t,0)$ are the evolution operators, free and in presence of the magnetic field, respectively.
Particles are assumed to enter and to exit the magnetic field region at times $t=0$ and $t=1,$ respectively.

Let us begin with the evolution in the magnetic field region.
Since $h_l=g_2\left(p_x^2 + p_z^2\right)$ and $h_m=g_1(x\sigma_1-z\sigma_3)$ do not commute, it is difficult to find an analytic expression for the unitary  operator $U_{xz} (t,0) = \exp \left[ -i \left( h_l + h_m \right) t \right]$; in fact, no closed expression is known.
However, the Suzuki-Trotter decomposition
\begin{equation}
U_{xz} (t,0)=\lim_{N_t\rightarrow\infty}\left[e^{-ih_m\frac{t}{2N_t}}e^{-ih_l\frac{t}{N_t}}e^{-ih_m\frac{t}{2N_t}}\right]^{N_t},
\label{eq:time_evolution_trotter}
\end{equation}
can be used as an approximation, by using a large but finite $N_t$.
This decomposition is used to iteratively find the state $|\psi(t)\rangle$,
 \begin{equation}
|\psi(t_n)\rangle=e^{-ih_m\frac{\delta t}{2}} e^{-ih_l\delta t} e^{-ih_m\frac{\delta t}{2}} |\psi(t_{n-1})\rangle,
\label{eq:state_trotter_iter}
\end{equation}
where $\delta t = t/N_t$, $|\psi(t_0)\rangle= |\psi_{xz}\rangle|\chi\rangle$, with $|\chi\rangle$ an arbitrary spin state, and $|\psi(t_{N_t})\rangle=|\psi(t)\rangle$.
The index $n$ runs from $0$ to $N_t$.

At each time step, three evolution operators are applied.
For the application of the first one, it is convenient to expand the state and the evolution operator  as $|\psi(t_n)\rangle = \sum_{\mu=0}^3|\phi_\mu (t_n)\rangle\sigma_\mu |\chi\rangle$ and $\exp\left(-ih_m \delta t/2 \right) = \sum_{\mu=0}^3 u _\mu(\delta t/ 2)\, \sigma_\mu,$ respectively. Hence,
\begin{align*}
e^{-ih_m\frac{\delta t}{2}} |\psi(t_{n})\rangle
& =\sum_{\mu = 0}^3 |\bar{\phi}_\mu(t_{n})\rangle \sigma_\mu |\chi\rangle,
\end{align*}
where
\begin{equation*}
\begin{gathered}
|\bar{\phi}_0(t_{n})\rangle = \sum_{\mu=0}^3 u_\mu (\delta t/2) |\phi_\mu (t_{n})\rangle,\\
\begin{aligned}
|\bar{\phi}_l (t_{n}) \rangle = & u_l (\delta t/2) |\phi_0(t_{n})\rangle + u_0 (\delta t/2) |\phi_l(t_n)\rangle\\
&+i\sum_{ij=1}^3 \varepsilon_{ijl}u_i (t)|\phi_j(t_n)\rangle , \quad l=1,2,3,
\end{aligned}
\end{gathered}    
\end{equation*}
and $\varepsilon_{ijl}$ is the completely anti-symmetric Levi-Civita symbol. 

The second operator to be applied at each time step is $\exp\left(-ih_lt/N_t\right)$. 
In this case, there is no need to expand the operator in terms of Pauli spin operators, because $h_l$ is defined only over $\mathcal{H}_{xz}$. 
However, since $h_l$ is multiplicative in the momentum representation, we must transform the state to this representation before applying the second unitary operator. The result is then transformed back to the position representation.
Without taking into account these transformations, we have 
$$e^{-ih_l\delta t} e^{-ih_m\frac{\delta t}{2}} |\psi (t_n)\rangle = \sum_{\mu=0}^3 |\tilde{\phi}_\mu(t_n)\rangle \sigma_\mu |\chi\rangle,$$
where $|\tilde{\phi}_\mu (t_n)\rangle = \exp\left(-ih_l \delta t\right) |\bar{\phi}_\mu (t_n)\rangle$.
The possible free evolution of the particles after they exit the magnetic field region, which has the same form, is handled in the same way.
At the end of each time step, we apply operator $\exp\left(-ih_mt/2N_t\right)$ once more. 
The procedure is exactly the same as in the first application. 

After $N_t$ steps and after considering the free evolution before detection, we obtain the approximated state $|\psi(T)\rangle\approx\sum_{\mu=0}^3|\phi_\mu(T)\rangle\sigma_\mu|\chi\rangle$. 
The vectors $|\phi_\mu(T)\rangle$ correspond to approximations to the vectors $A_\mu(T)|\psi_{xz}\rangle$, so the values $\langle x,z|\phi_{\mu}(T)\rangle$ correspond to approximations to the functions $\phi_\mu(x,z,T)$.

The implementation of the previous method requires the additional step of discretizing both position and momentum. The $x$ and $z$ coordinates are sampled over the intervals $\left[x_{\textrm{min}},x_{\textrm{max}}\right]$ and $\left[z_{\textrm{min}},z_{\textrm{max}}\right]$, at sampling frequencies $\delta x = \left(x_{\textrm{max}}-x_{\textrm{min}}\right)/N_x$ and $\delta z=\left(z_{\textrm{max}}-z_{\textrm{min}}\right)/N_z$, respectively. Here, $N_x$ and $N_z$ indicate the number of samples in each coordinate. It is important that the coordinate intervals are large enough to reduce the effects generated by the artificial boundary conditions~\cite{Hsu2011}. Similarly, the momentum coordinates $p_x$ and $p_z$ are sampled in steps of $\delta p_x=2\pi/(N_x\delta x) $ and $\delta p_z=2\pi/(N_z\delta z) $, over the intervals $\left[-\pi/\delta x, \pi/\delta x \right]$ and $\left[-\pi/\delta z, \pi/\delta z \right]$. 
Vectors $|\phi_\mu(t_n)\rangle$ and operators $u_\mu (\delta t/2)$ and $\exp\left(-ih_l \delta t\right)$ then turn into $(N_x+1)\times (N_z+1)$ arrays. Accordingly, the multiplicative application of discretized operators over discretized states becomes a Hadamard (element-wise) product between arrays of the same size and the transformation from position to momentum representation becomes a fast Fourier transform. 

Having found the arrays $\phi_\mu (x,z,T)$, functions $M_\mu(x,z,T)$ are calculated form relation~\eqref{eq:func_comp_exp}. For the elements $M_{k\mu}$, the integral in equation~\eqref{eq:matrix_comp_exp} is computed as
\begin{equation*}
\sum_{(x_i,z_j)\in \Omega_k} \phi_\alpha(x_i,z_j,T)
\phi_\beta^{*}(x_i,z_j,T)\delta x\delta z,
\end{equation*}
where the sum extends over the pairs $(x_i,z_j)$ belonging to the region $\Omega_k.$

\nocite{*}
\bibliography{apssamp}

\end{document}